%% file: jvla_spectralageing.tex
\documentclass{mn2e} 
\usepackage{epsf,times}
\usepackage{amsmath}
\usepackage{graphicx}
\usepackage{color}
\usepackage{natbib}
\usepackage{times}

\usepackage{comment}

\newcommand{\etal}{{et al}\/.}

\begin{document}

\title[Spectral ageing in cluster-centre FR-IIs]{Spectral ageing in the lobes of cluster-centre FR-II radio galaxies}

\author[J.J.~Harwood \etal]{Jeremy J.\ Harwood$^{1}$\thanks{E-mail: jeremy.harwood@physics.org}, Martin J.\ Hardcastle$^{2}$ and Judith H.\ Croston$^{3,4}$ 
\\$^{1}$ASTRON, The Netherlands Institute for Radio Astronomy, Postbus 2, 7990 AA, Dwingeloo, The Netherlands
\\$^{2}$School of Physics, Astronomy and Mathematics, University of Hertfordshire, College Lane, Hatfield, Hertfordshire AL10 9AB, UK
\\$^{3}$School of Physics and Astronomy, University of Southampton, Southampton SO17 1BJ, UK,
\\$^{4}$Institute of Continuing Education, University of Cambridge, Madingley Hall, Madingley, CB23 8AQ, UK
}

\maketitle


\graphicspath{{./images/}}

\input{./abstract.tex}

\input{./keywords.tex}

\input{./introduction.tex}

\input{./method.tex}

\input{./results.tex}

\input{./discussion.tex}

\input{./conclusions.tex}
\input{./acknowledgements.tex}

\bibliographystyle{mn2e}

\def\newblock{\hskip .11em plus .33em minus .07em}
\bibliography{jvla_spectralageing}

\input{./appendix.tex}

\end{document}

%% file: abstract.tex
\begin{abstract}

Recent investigations have shown that many parameters and assumptions made in the application of spectral ageing models to FR-II radio galaxies (e.g. injection index, uniform magnetic field, non-negligible cross-lobe age variations) may not be as reliable as previously thought. In this paper we use new VLA observations, which allow spectral curvature at GHz frequencies to be determined in much greater detail than has previously been possible, to investigate two cluster-centre radio galaxies, 3C438 and 3C28. We find that for both sources the injection index is much steeper than the values traditionally assumed, consistent with our previous findings. We suggest that the Tribble model of spectral ageing provides the most convincing description when both goodness-of-fit and physically plausibility are considered, but show that even with greatly improved coverage at GHz frequencies, a disparity exists in cluster-centre FR-IIs when spectral ages are compared to those determined from a dynamical viewpoint. We find for 3C438 that although the observations indicate the lobes are expanding, its energetics suggest that the radiating particles and magnetic field at equipartition cannot provide the necessary pressure to support the lobes, similar to other cluster-centre source such as Cygnus A. We confirm that small scale, cross-lobe age variations are likely to be common in FR-II sources and should be properly accounted for when undertaking spectral ageing studies. Contrary to the assumption of some previous studies, we also show that 3C28 is an FR-II (rather than FR-I) source, and suggest that it is most likely a relic system with the central engine being turned off between 6 and 9 Myrs ago.

\end{abstract}

%% file: keywords.tex

%% file: introduction.tex
\begin{table*}
\centering
\caption{List of target sources and galaxy properties}
\label{jvlatargets}
\begin{tabular}{llcccccc}
\hline
\hline

Name&IAU Name&Redshift&178 MHz&5 GHz Core&Spectral Index&LAS&Size\\
&&&Flux (Jy)&Flux (mJy)&(178 to 750 MHz)&(arcsec)&(kpc)\\
\hline
3C438&J2153$+$377&0.290&48.7&7.1&0.88&22.6&98.3\\
3C28&J0053$+$261&0.195&17.8&$<0.2$&1.06&45.6&148\\
\hline

\end{tabular}

\vskip 5pt
\begin{minipage}{12.3cm}
`Name' and `IAU Name' list the 3C and IAU names of the galaxies as discussed in this paper. `Spectral Index' lists the low frequency spectral index between 178 to 750 MHz, `LAS' the largest angular size of the source and `Size' its largest physical size. The `Redshift', `178 MHz Flux', `5 GHz Core Flux', `Spectral Index' , `LAS' and `Size' column values are taken directly from the 3CRR database \citep{laing83} (http://3crr.extragalactic.info/cgi/database).
\end{minipage}
\end{table*}

\section{Introduction}
\label{intro}

\subsection{Radio galaxies}
\label{specageintro}

Powerful \citet{fanaroff74} class I (FR-I) and II (FR-II) radio galaxies can have a significant impact on the environment in which they reside, with structure that can extend from tens of kiloparsecs \citep{birkinshaw81, alexander87, konar06, machalski09} to megaparsecs \citep{mullin06, marscher11} in size. The ability of these radio loud active galaxies to provide the required energy input to suppress star formation in models of galaxy evolution through AGN feedback \citep{croton06, bower06} has therefore proved popular in explaining the observed properties of stellar populations in massive ellipticals. However, many aspects of the underlying dynamics and energetics of these powerful radio sources remain a mystery.

FR-II radio galaxies generally consist of three large scale structures: jets, lobes and hotspots. The jets of FR-IIs are relativistic and are thought to be the mechanism for transporting material from the central active galactic nucleus (AGN) to a termination shock, usually located at the extremities of the source. This shock forms a compact region of synchrotron emission, known as the hotspot, and is generally thought to be the dominant location of particle acceleration within these sources. As the hotspot moves away from the nucleus \citep{burch77, burch79, winter80, meisenheimer89}, the previously accelerated particles are left behind (possibly with some back flow of material) but continue to radiate via the synchrotron process which are then observed as the lobes which give FR-IIs their characteristic morphology (e.g. \citealp{scheuer74, begelman89, kaiser97, krause12}). As these lobes are in direct contact with the external medium, if we are to determine how these powerful radio sources affect the evolution of galaxies as a whole, then we must understand their dynamics, energetics and ultimately the total energy they transfer to their environment.

\subsection{Spectral ageing}
\label{specagemodels}

In theory, for an electron which is emitting via the synchrotron process in a fixed magnetic field, the energy losses scale as \begin{equation}\label{elosses}\tau = \frac{E}{dE/dt} \propto 1/E \propto 1/\nu^{2} \end{equation} This leads to a preferential cooling of higher energy electrons and, in the absence of any further particle acceleration, produces a spectrum which becomes increasingly curved over time. Thus, for electron energy distribution initially described by a power law \begin{equation}\label{initialpowerlaw}N(E) = N_0 E^{-\delta}\end{equation} we find at later times that \begin{equation}\label{edistribution} N(E,\theta,t) = N_0 E^{-\delta} (1 - E_{T} E)^{-\delta - 2}\end{equation} where $E_{T}$ are the model dependent losses which are a function of the pitch angle of the electrons to the magnetic field $\theta$, and time since acceleration $t$ (see \citealp{harwood13} for a detailed discussion). This time dependent process, known as a spectral ageing, has become a common method for determining the age of sources radiating via the synchrotron process.

The ability of spectral ageing models to describe the emission from the lobes of radio galaxies, particularly their ability to provide the intrinsic age of a source, has long been a topic of debate (e.g. \citealp{alexander87, eilek96, eilek97, blundell00}); however, the technique remains a commonly used tool in the analysis of both low- and high-power radio galaxies (e.g. \citealp{jamrozy05, kharb08, orru10, degasperin12, heesen14}). Work to develop more advanced ageing models that better describe the observed emission continues to this day \citep{tribble93, komissarov94, machalski09, hardcastle13a} but testing of spectral ageing models against the new generation of radio interferometers that allow us to obtain much tighter constraints remains largely unexplored.

Our previous work \citep{harwood13} has given the first insights into the spectral structure of FR-II radio galaxies on small spatial scales and provided the methods required to allow the detailed spectral study of these sources now that truly broad-bandwidth observations are available. Whilst this investigation found that spectral ageing has at least some basis in reality for powerful radio sources, it has also become apparent that many of the previously held assumptions made in the application of models of spectral ageing are, at least in some cases, less reliable than previously thought.

The first of these potentially incorrect assumptions was that the model parameter which describes the observed spectrum of the initial power law electron energy distribution (the injection index) is much steeper than previously assumed. This injection index, which is directly related to $\delta$ by \begin{equation}\label{alphainject}\alpha_{inj} = \frac{\delta - 1}{2}\end{equation} is traditionally assumed to have values of around 0.5 to 0.6 based on theoretical arguments (e.g. \citealp{blandford78}) and observations of hotspots (e.g. \citealp{meisenheimer89, carilli91}). However, our previous investigation suggests that, for the two sources studied, the injection index is $>$0.8 in both cases. The increased low-frequency energy content that this implies has implications for both the spectral ages of powerful radio galaxies and their energetics, so it is key that we determine if these findings are robust and if they are common to the FR-II population as a whole.

The second issue, first discussed by \citet{eilek96a}, was a disparity between the spectral ages and those determined from a dynamical view point. One potential solution to this problem was to place tighter constraints on the curvature of the lobe spectrum; however, our study showed that even with greatly improved sampling in frequency space, this disparity still remains. In order to resolve this issue, it is important to determine whether this age difference is found in all galaxies, or if it is confined to FR-IIs in certain environments.

The final problem directly related to the work described within this paper is determining which model of spectral ageing best describes the spectrum of the sources, both in terms of goodness-of-fit and in their physical interpretation. We discuss the various models of spectral ageing further in Section \ref{modelfitting} (also see \citealp{harwood13}) but note here that the most commonly applied model of spectral ageing proposed by \citet{jaffe73} (the JP model) is frequently in conflict with that of \citet{kardashev62} and \citet{pacholczyk70} (the KP model). The JP model is often preferred due to its physical plausibility but often provides a worse goodness-of-fit than the less physically realistic KP model, and is what we also find in our 2013 study. We therefore tested the more recent Tribble model of spectral ageing \citep{tribble93, hardcastle13a, harwood13} which accounts for a more complex description of the magnetic field within the lobes. We found that this model was both able to provide a comparable goodness-of-fit to the KP model and to retain the physical plausibility of the JP model. If one is to determine the impact of powerful radio galaxies on their environment then resolving these outstanding issues and testing these new models in a range of environments in order to be able to reliably determine a sources age, hence total power output over its lifetime, is a vital step.

\begin{figure*}
\centering
\includegraphics[angle=0,height=6.3cm]{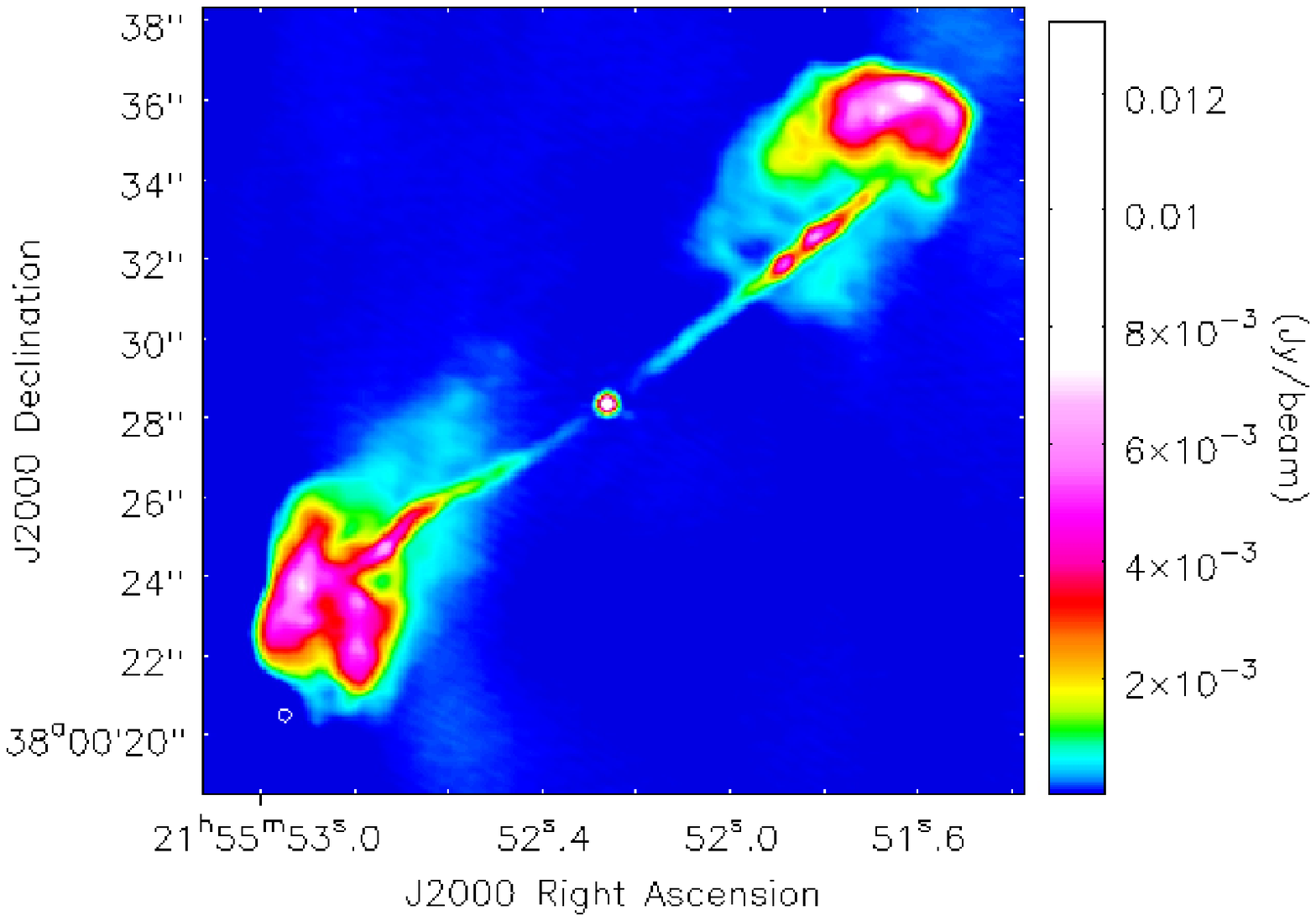}
\includegraphics[angle=0,height=6.3cm]{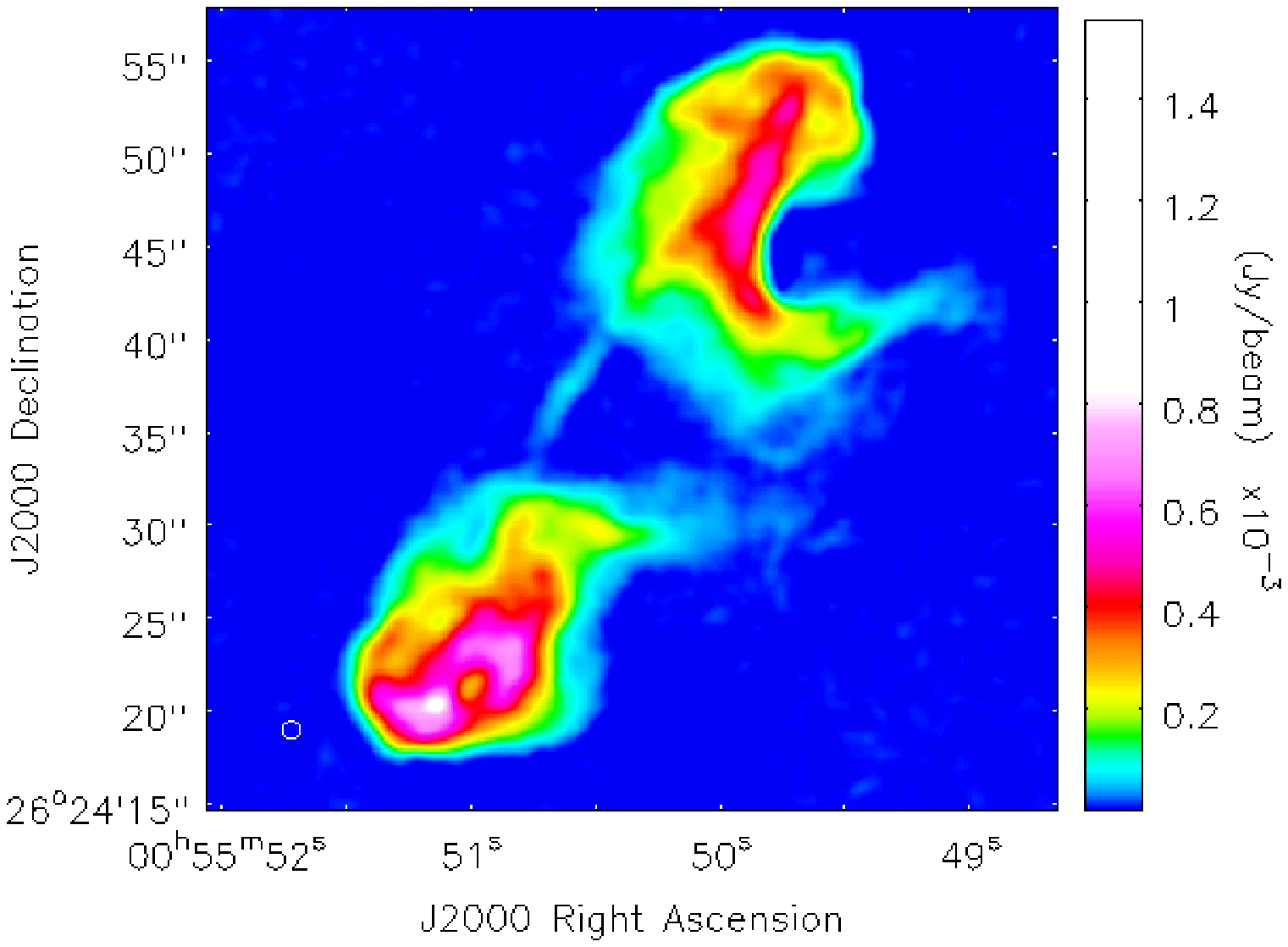}\\
\caption{Combined A, B and C configuration radio map of 3C438 (left) and B and C configuration radio map of 3C28 (right) between 4 and 8 GHz. Imaged using multiscale CLEAN and \textsc{casa} nterms = 2 to a central frequency of 6 GHz (see Section \ref{datareduction} for details). The off-source RMS of the combined maps is $8 \mu$Jy beam$^{-1}$ and $4 \mu$Jy beam$^{-1}$ for 3C438 and 3C28 respectively. The restoring beams are 0.30 and 0.99 arcsec for 3C438 and 3C28 respectively and are indicated in the bottom left corner of each image.}
\label{combinedimages}
\end{figure*}

\subsection{Outstanding questions}
\label{outstandingquestions}

In order to determine new, reliable parameters and models of spectral ageing, we must first investigate whether our previous findings \citep{harwood13} are intrinsic to a wide range of FR-II sources in differing environments, a result of being still too poorly constrained in frequency space, or simply outliers in an otherwise expected distribution. Within this paper we therefore expand upon our initial sample using truly broad bandwidth observations of two cluster-centre radio galaxies to address 4 primary questions:

\begin{enumerate}
\item What are the dynamics and energetics of FR-II sources when tightly constrained by broad bandwidth data?\\
\item Which model of spectral ageing provides the best description of cluster-centre FR-IIs?\\
\item Is the injection index of cluster-centre FR-IIs flat, as had previously been assumed, or steep?\\
\item Does the well known dynamical versus spectral age disparity still exist in cluster-core FR-IIs?
\end{enumerate}

Section \ref{method} gives details of target selection, data reduction and the spectral analysis methods used. In Section \ref{results} we present our results and Section \ref{discussion} discusses the implications for our current understanding of FR-II radio galaxies. Throughout this paper, a concordance model in which $H_0=71$ km s$^{-1}$ Mpc$^{-1}$, $\Omega _m =0.27$ and $\Omega _\Lambda =0.73$ is used \citep{spergel03}.

%% file: method.tex
\section{Data Reduction and Spectral Analysis}
\label{method}

\subsection{Target selection and observations}

As curvature due to spectral ageing becomes most easily observable above a few GHz (\citealp{alexander87}; \citealp{alexander87a}; \citealp{carilli91}; \citealp*{perley97}; \citealp{hardcastle01}), the resolution obtained with the Karl G. Jansky Very Large Array (VLA) at C-band frequencies (4.0-8.0 GHz) made it the natural choice for the required observations. Using the 3CRR sample \citep*{laing83} of well-studied FR-II sources as reference, we selected our targets so as to ensure that they were both well sampled by the shortest baselines at the highest C-band frequencies, while still maintaining a reasonable resolution at the lowest frequencies. We therefore restricted the largest angular size (LAS) of the sources on the sky to $LAS \leq 50$ arcsec at a redshift of $z < 0.5$. We also imposed a spectral index cut-off for the integrated spectrum of $\alpha > 0.8$, where $0.8$ is the average of the 3CRR sample. This ensures that the targets are likely to show significant spectral ageing rather than being very young, nearby sources, selected due to the small apparent size of the LAS cut.

From the eight sources that satisfied these criteria in 3CRR, we excluded the peculiar sources 3C171 and 3C305, where the radio structure plausibly results from strong interaction with cold gas within the host galaxy \citep{hardcastle10}, and 4C14.27 which is likely to be a relic system. 17.5 hours of observing time was awarded which was suitable for observation of 2 of the remaining 5 sources in the required configurations. We therefore split the time equally between 3C438 and 3C28, both of which reside in cluster-centre environments. Details of the general properties of these sources are given in Table \ref{jvlatargets}.

In order to ensure that all known compact and diffuse emission was sampled, observations were made using the A, B and C array configurations. As 3C28 lacks any compact structure on scales that are sampled by the A configuration array, only B and C were used in the final images. At the time of observation the maximum bandwidth capability of the VLA was 2 GHz as the 3-bit samplers were not available; we therefore used two 2 GHz observations to give full frequency coverage between 4 and 8 GHz.

\subsection{Data reduction}
\label{datareduction}

The data were downloaded from the NRAO archive and reduced using {\sc casa} in the standard manner as described in the \textsc{casa} cookbook\footnote{http://casa.nrao.edu/docs/UserMan/UserMan.html} and the online tutorial for narrow-field, C-band observations \footnote{http://casaguides.nrao.edu/index.php?title=EVLA\_6cmWideband\_Tutorial\\\_SN2010FZ}. As multiple images were required for our analysis rather than just a single broadband data cube, each of the 32 125-MHz spectral windows was individually calibrated.

The structures of 3C438 and 3C28 differ significantly, hence different imaging strategies were required to obtain the optimal radio maps. For 3C438, the A-configuration data were first self-calibrated in phase only and CLEANed to convergence. The A configuration images were then used as a model to cross-calibrate the B-configuration observations. The A and B configuration \emph{uv} data were then combined and imaged, and these images in turn were used to cross-calibrate the C configuration data. As 3C28 does not contain compact structure, it was not possible to apply the same the cross-calibration method used for 3C438. Instead, the B and C configuration \emph{uv} data were combined and a standard self-calibration performed in phase only. The radio maps of both sources were produced using the multi-scale clean algorithm MSCLEAN \citep{cornwell08, rau11} at scales of 0, 5, 15 and 45 times the cell size, where the cell size was set to one fifth of the beam size. In order for a detailed spectral analysis to be performed the parameters of each image must be matched, so the beam size of the final images was set to that of the lowest frequency data. We excluded the two lowest spectral windows for both sources (for reasons discussed in Section \ref{dataquality}) and so image at the resolution of the lowest frequency used in the analysis, rather than the lowest frequency observed. A summary of parameters used for the imaging described in this section is shown in Table \ref{cleanparams}.

As the observations described so far were taken for each array configuration in a single pointing we can be confident that they are inherently well aligned. However in order to correct for small, sub pixel variations in position, Gaussians were fitted to the radio core for 3C438 and to a point source close to 3C28 (where the radio core is not clearly visible in all images) using \textsc{casaviewer}. A reference pixel was then decided upon and each map aligned using the \textsc{aips}\footnote{http://www.aips.nrao.edu/index.shtml} OGEOM task.

In order to extend our frequency coverage the combined VLA / MERLIN L-band data of 3C438 presented by \citet{treichel01} and data for 3C28 from the atlas of DRAGNs (Leahy, Bridle \& Strom\footnote{http://www.jb.man.ac.uk/atlas/}), which are of comparable resolution to the C-band observations, were also obtained. These data were already calibrated to a high standard, hence only regridding to J2000 coordinates, re-imaging to match the parameters of Table \ref{cleanparams} and alignment as described above was required.

To check the impact of any residual alignment errors, Gaussians were again fitted to the aligned maps. We find the standard deviation in the alignment between maps to be only 0.007 pixels, and so misalignments are unlikely to make any significant contribution to the uncertainty in the fitting of models to the oldest regions of plasma.

Additional images were also created using the full bandwidth centred at a frequency of 6 GHz. As, for the combined data set, the flux changes significantly over the bandwidth, the data must be suitably scaled in order to produce a realistic image of the source. We therefore used the \textsc{casa} multi-frequency synthesis (MFS) CLEAN parameter $nterms=2$ \citep{rau11} which scales the flux by a spectral index value fitted over the observed bandwidth. The resulting images are shown in Figure \ref{combinedimages}.

\begin{table}
\centering
\caption{Summary of imaging parameters}
\label{cleanparams}
\begin{tabular}{lllcl}
\hline
\hline
Source&Parameter&\textsc{casa} Name&Value&Units\\
\hline
3C438&Polarization&\textsc{stokes}&I&\\
&Image Size&\textsc{imsize}&4096 4096&Pixels\\
&Weighting&\textsc{robust}&-0.5&\\
&Cell Size&\textsc{cell}&0.084 0.084&Arcsec\\
&Beam size&\textsc{restoringbeam}&0.42 0.42&Arcsec\\
&Multiscale&\textsc{multiscale}&[0, 5, 15, 45]&Pixels\\
3C28&Polarization&\textsc{stokes}&I&\\
&Image Size&\textsc{imsize}&6144 6144&Pixels\\
&Weighting&\textsc{robust}&1&\\
&Cell Size&\textsc{cell}&0.42 0.42&Arcsec\\
&Beam size&\textsc{restoringbeam}&2.1 2.1&Arcsec\\
&Multiscale&\textsc{multiscale}&[0, 5, 15, 45]&Pixels\\
\hline
\end{tabular}
\vskip 5pt
\begin{minipage}{8.5cm}
`Parameter' refers to the imaging parameter used in making of radio maps within this chapter. `{\sc casa} Name' refers to the {\sc casa} parameter taking the value stated in the `Values' column.
\end{minipage}
\end{table}

\subsection{Data quality}
\label{dataquality}

From the 32 VLA spectral windows between 4 and 8 GHz, we found that 28 had good quality data in all array configurations for both 3C438 and 3C28. Of the remaining four, the spectral window at 6.10 GHz (SPW 17) was of extremely poor quality in all configurations and was therefore excluded from our analysis. At 4.02 GHz (SPW 0), heavy flagging of RFI was required in all array configurations leading to poor data quality and so this image was excluded from further analysis. The observations at 4.14 and 6.36 GHz (SPW 1 and 19 respectively) are of reasonable quality, but significant RFI was present in the B and C configuration data. These two frequencies were therefore excluded from the spectral analysis, but included in the making of the combined image shown in Figure \ref{combinedimages}. As these problems are common to both sources, we suggest that they are most likely due to an instrumental issue at the time of observation. It is also worth noting that the low flux density `hole' observed in the northern lobe of 3C438 is also seen in the atlas of DRAGNs and in observations made by \citet{hardcastle97} and so is likely to be physical in origin, rather than an imaging artefact.

Within the field of view of our observations of 3C28 another large, bright, extended radio source was also serendipitously observed. Unfortunately, due to being located at the edge of the JVLA's primary beam and with no well matched L-band data available, we are unable to perform a full spectral analysis. However, we are still able to provide some insight into the source's properties and therefore provide a brief analysis of this peculiar source in Appendix \ref{nat}.

\subsection{Spectral analysis}
\label{spectralanalysis}

For the analysis of our sources we have used the Broadband Radio Astronomy Tools\footnote{http://www.askanastronomer.co.uk/brats} (\textsc{brats}) software package which provides a wide range of spectral analysis and spectral age model fitting tools. The basic usage and functionality of \textsc{brats} has been discussed in detail previously \citep{harwood13} and in the \textsc{brats} cookbook\footnote{http://www.askanastronomer.co.uk/brats/bratscookbook.pdf} and so we do not provide an in depth discussion here. We instead give a brief summary of the fitting performed and highlight new features which have been added to the software and used within this paper over previous works.

\subsubsection{Region selection}
\label{regionselection}

A region fully encompassing the source (but excluding the core) was defined using \textsc{ds9}\footnote{http://ds9.si.edu}, along with a background region for determining the off-source thermal noise. To test the impact of the strong emission in 3C438 coincident with the assumed location of the jet on the model parameters and fits, a second region was also created with this area excluded. Initial source detection was then performed, for which a $5 \sigma$ cut-off was used based on this RMS value. We adopt the same practice used by \citet{harwood13} and apply an on-source multiplier of 3 times this value for the determination of statistical values and region selection to account for the increased RMS noise due to the increased uncertainty in the modelling of the extended emission. As each VLA image is taken within a single pointing, one can be confident that the relative flux calibration error between images is small. We therefore take the flux calibration error to be 1 per cent for the VLA images between 4 and 8 GHz. As the L-band observations were made with MERLIN and the old VLA system, the flux calibration error relative to the VLA is not immediately clear. We therefore take a conservative value of the standard absolute flux calibration error value for VLA L-band observations of 5 per cent. To reduce the impact of the superposition of spectra on our results \citep{harwood13, stroe14}, and as these data are of good quality, in the analysis that follows we considered each source on a pixel by pixel basis. A summary of the values chosen for the region selection is shown in Table \ref{regprms}.

\subsubsection{Model fitting and parameter determination}
\label{modelfitting}

We have tested the broad bandwidth data described above against three single-injection models of spectral ageing. The two most common models used to describe emission from the lobes of FR-II galaxies are those proposed by \citet{kardashev62} and \citet{pacholczyk70} (the KP model) and by \citet{jaffe73} (the JP model). These models, which have been widely discussed elsewhere (e.g. \citealp{leahy91, blundell01, harwood13}), differ only in their treatment of the pitch angle of electrons with respect to the magnetic field. The KP model assumes the angle of any given electron is fixed over its radiative lifetime, compared to the JP model which uses a time averaged value. This time averaging leads to an exponential cut-off at high frequencies, compared to the KP model which is comparatively flat due to there being a supply of high energy electrons at small pitch angles which are capable of radiating at higher frequencies. 

The third model tested is that proposed by \citet{tribble93} which attempts to account for a spatially non-uniform magnetic field. Both the KP and JP models assume a fixed magnetic field strength but this is unlikely to be physically realistic. The Tribble models therefore attempt to account for the magnetic field structure by assuming it to be Gaussian random and allowing electrons to diffuse across regions of varying field strength. \citet{hardcastle13a} has recently shown that in the weak field, high diffusion case where electrons are free-streaming, the spectrum can be modelled by integrating the standard JP losses over a Maxwell-Boltzmann distribution in field energy density. We have previously detailed the application of this model to similar FR-II sources \citep{harwood13} suggesting that it may provide a solution to an outstanding problem where in many cases the KP model provides a better goodness-of-fit than the more physically realistic JP model.

We have made the standard assumption that the magnetic field (or in the case of the Tribble model, mean magnetic field) is in equipartition, which we calculate using the {\sc synch} code of \citet*{hardcastle98}. We took the minimum and maximum electron Lorentz factor to be $\gamma = 10$ and $\gamma = 1 \times 10^{6}$ respectively. To determine the best fitting injection index we have used \textsc{brats}' `\emph{findinject}' command between $\alpha_{inj} = 0.5$ and $1.0$ for 3C438 and between $0.5$ and $1.5$ for 3C28 at intervals of $0.1$. A second run was then performed around the minimum with a smaller step size of $0.01$. The resulting values were then plotted and the minimum injection index found. Final model fitting of the sources was then performed in \textsc{brats} using the derived values. The statistical values for each model were recorded and spectral ageing, $\chi^{2}$ and error maps were then exported ready for further analysis.

\begin{table}
\centering
\caption{Summary of image noise by frequency}
\label{mapdetails}
\begin{tabular}{ccccccc}
\hline
\hline

&\multicolumn{2}{c}{3C438}&\multicolumn{2}{c}{3C28}\\
Frequency&Off-source&On-source&Off-source&On-source\\
(GHz)&($\mu$Jy/beam)&($\mu$Jy/beam)&($\mu$Jy/beam)&($\mu$Jy/beam)\\

\hline
1.42&70.6&212&-&-\\
1.46&-&-&75.3&226\\
4.33&34.1&102&16.6&49.8\\
4.45&33.5&101&15.8&47.4\\
4.58&35.6&107&16.3&48.9\\
4.71&34.1&102&16.2&48.6\\
4.84&33.8&101&14.4&43.2\\
4.97&32.1&96.4&14.4&43.2\\
5.07&38.8&116&15.5&46.5\\
5.20&33.5&100&14.2&42.6\\
5.33&32.2&101&14.7&44.1\\
5.45&30.9&96.7&15.1&45.3\\
5.58&28.8&92.7&15.0&45.0\\
5.71&28.7&86.4&14.6&43.8\\
5.84&28.7&86.0&15.4&46.2\\
5.97&31.6&94.8&16.2&48.6\\
6.03&36.0&108&11.9&35.7\\
6.29&25.8&77.3&9.77&29.3\\
6.54&25.7&77.0&9.63&28.9\\
6.67&27.6&82.9&10.6&31.8\\
6.80&25.5&77.6&9.39&28.2\\
6.93&26.3&79.0&9.40&28.2\\
7.03&29.6&88.8&10.2&30.6\\
7.16&23.9&71.7&9.12&27.4\\
7.29&23.8&71.3&9.31&27.9\\
7.41&23.9&71.8&9.52&28.6\\
7.54&25.7&77.1&9.85&29.6\\
7.67&24.8&74.5&9.54&28.6\\
7.80&23.8&71.4&9.76&29.3\\
7.93&24.7&74.2&9.45&28.4\\
\hline

\end{tabular}

\vskip 5pt
\begin{minipage}{8.5cm}
`Frequency' refers to the frequency of the map, `Off-source RMS' refers to RMS noise measured over a large region well away from the source and `On-source RMS' the noise used for region selection and statistics as per Section \ref{spectralanalysis}.
\end{minipage}

\end{table}

\subsubsection{Uncertainty calculation}
\label{errors}

One key feature which has been applied to the data over previous investigations is the inclusion of improved uncertainty calculations. During the fitting process \textsc{brats} performs a search over a large range of potential spectral ages allowing the $\chi^{2}$ curve as a function of age to be determined. \citet{avni76} shows that the $1\sigma$ error of such a minimization is given by a deviation of $\Delta \chi^{2} = 1$ from the minimum $\chi^{2}$ value. These errors were therefore calculated during the fitting process, mapped as a function of position, and then exported for further analysis. A similar error calculation was also performed to provide $1\sigma$ errors for the injection index values. However, as here one uses the sum of $\chi^{2}$ over all regions for each injection index to determine the minimum value, they are over weighted by a factor of the beam area. Assuming the injection index is approximately constant across any given source, the injection index errors are therefore determined by finding where $\Delta \chi^{2} = 1$ for the corrected $\chi^{2}$  values such that $\chi^{2}_{corr} = \chi^{2} / A_{beam}$ where $A_{beam}$ is the beam area. These methods provide a much more robust error estimate than in previous studies and are applied throughout this paper.

\begin{table}
\centering
\caption{Summary of adaptive region parameters}
\label{regprms}
\begin{tabular}{lcll}
\hline
\hline

Parameter&Value&Units&Description\\

\hline
Signal to noise&1&&SNR (pixel to pixel)\\
Search area&1&Pixels$^{2}$&Max. search area\\
On-source noise&3&&On-source noise multiplier\\
Hot pixel limit&20&Per cent&Max. pixel variation\\
Map variations&-1&&Maximum map variation (off)\\
\hline

\end{tabular}

\vskip 5pt
\begin{minipage}{8.5cm}
`Value' refers to the values applied within {\sc brats} for the corresponding `Parameter'. The `Description' column provides further details of the value meaning. Note that a signal to noise of 1 is still subject to the 5$\sigma$ cutoff and the on-source noise multiplier.
\end{minipage}

\end{table}

\subsubsection{Spectral index fitting}
\label{specindexfitting}

An additional improvement to \textsc{brats} is the enhanced handling of spectral index fits. As the uncertainty on the flux measurements of radio observations varies between images, observations and telescopes, the assumption of constant errors used by a standard least-squares fit is often not valid. \textsc{brats} therefore now includes a weighted least-squares option for spectral index determination. This function fits a linear regression in log space of the standard form $y(c,x)=c_{0} + c_{1} x$ but, instead of assuming uniform errors, uses the GSL `wlinear' and `linear\_est' functions\footnote{https://www.gnu.org/software/gsl/manual/html\_node/Linear-regression.html\#Linear-regression} to minimize the weighted sum of squared residuals such that when fitting to $n$ observations \begin{equation}\label{weightedlsf}\chi^{2} = \sum_{i=1}^{n} w_{i}(y_{i} - (c_{0} + c_{1} x_{i}) )^{2} \end{equation} where the weights are given by $w = 1/\sigma_{i}^{2}$ and $\sigma_{i}$ is the error on a given flux measurement. Note that as fitting is performed in log-log space, the fractional errors on the measurements must be small so that one can assume they remain approximately symmetric when transformed to natural log space. In general, this assumption is valid for fractional errors $<$10 per cent and so is well within the uncertainties used within this paper.

\begin{figure*}
\centering
\includegraphics[angle=270,width=8.8cm]{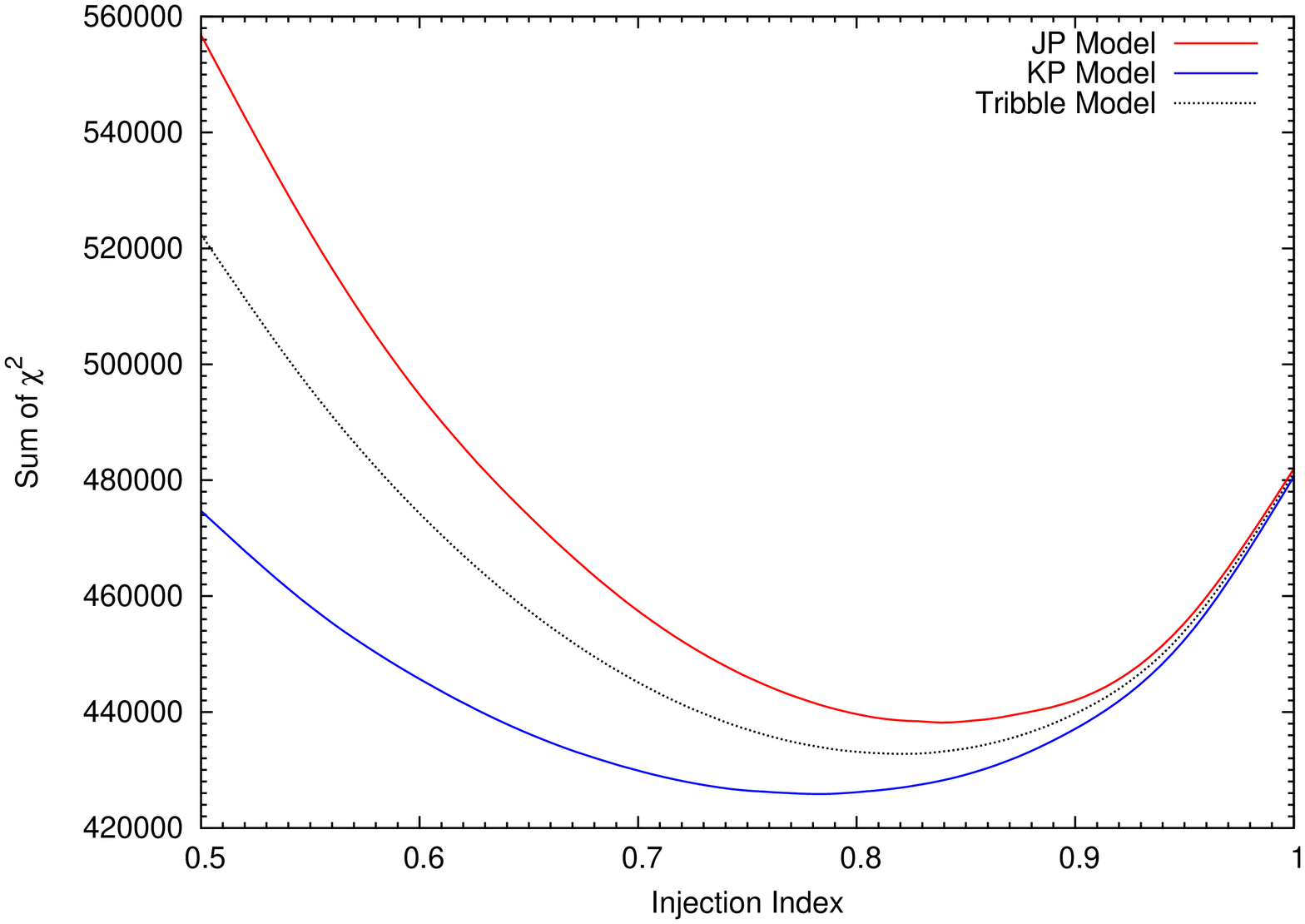}
\includegraphics[angle=270,width=8.8cm]{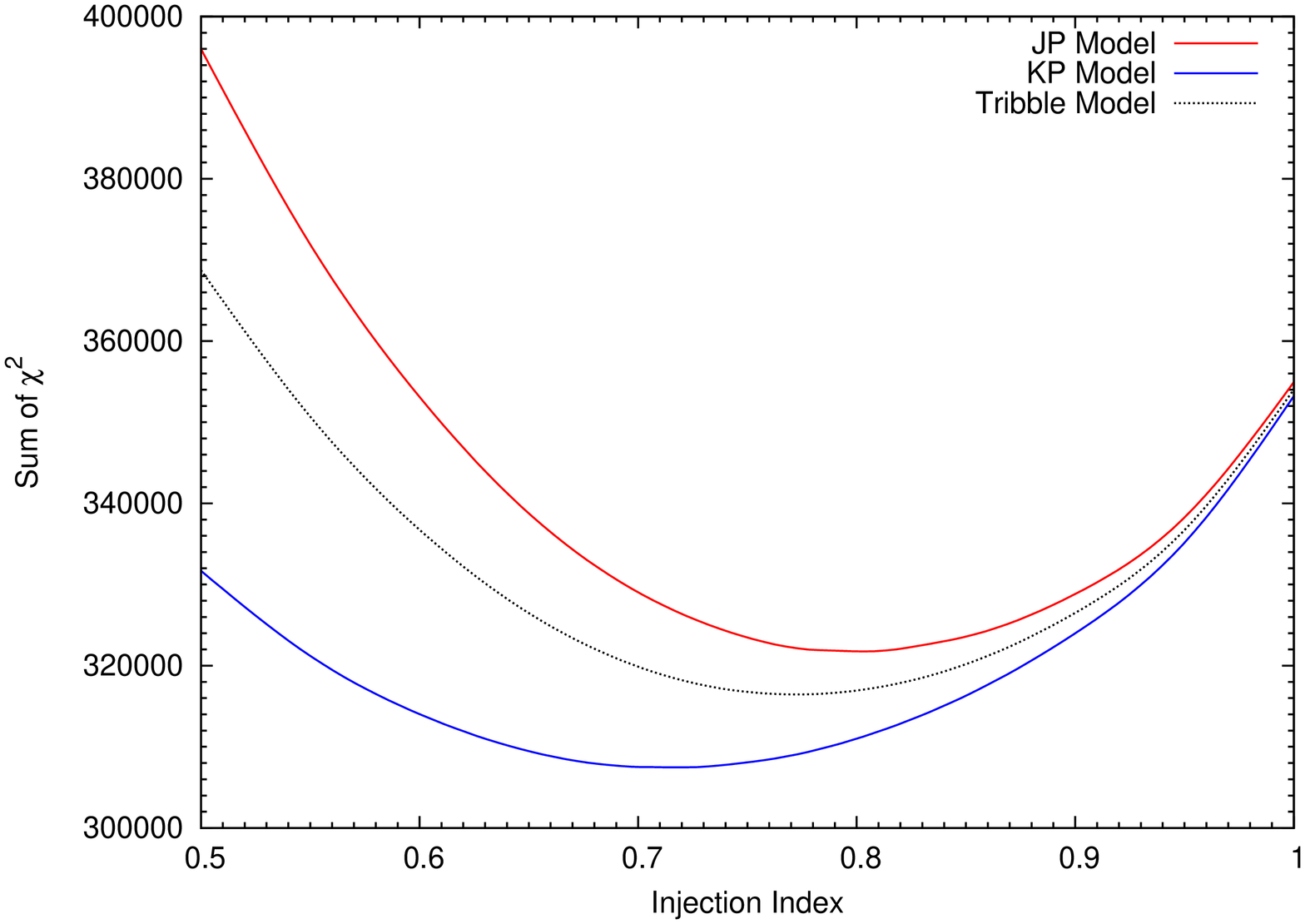}\\
\includegraphics[angle=270,width=8.8cm]{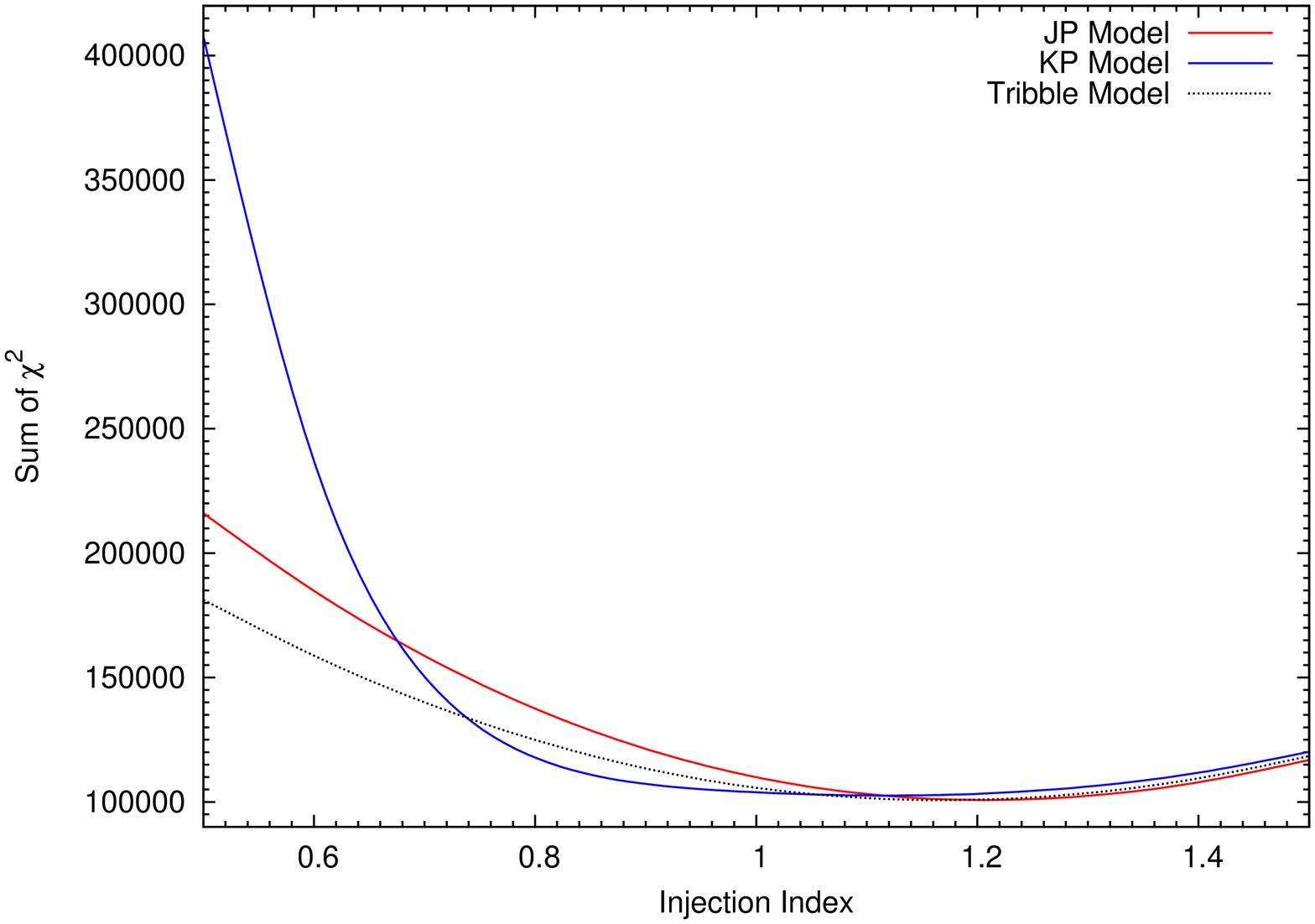}\\
\caption{$\chi^{2}$ values for 3C438 including the jet emission (top left), excluding the jet emission (top right), and 3C28 (bottom) for varying injection index values using the JP, KP and Tribble models of spectral ageing. Data points are taken at intervals of 0.1 between 0.5 and 1.0 for 3C438 and 0.5 and 1.5 for 3C28, with intervals of 0.01 around the minimum values. The data are fitted with natural cubic splines. As all points lie on the fitted spline they are excluded for clarity. The best fitting injection index values are shown in Table \ref{bestinject}.}
\label{jvlainjectmin}
\end{figure*}

%% file: results.tex
\section{Results}
\label{results}

\begin{table}
\centering
\caption{Best fitting injection indices and magnetic field strengths}
\label{bestinject}
\begin{tabular}{llcccc}
\hline
\hline

Source&Model&Injection&\multicolumn{2}{c}{Error}&Magnetic Field\\
&&Index&+&-&Strength (nT)\\

\hline
3C438&JP&0.84&0.01&0.01&4.02\\
&KP&0.78&0.01&0.01&3.65\\
&Tribble&0.82&0.01&0.01&3.89\\
3C438&JP&0.80&0.01&0.01&3.77\\
(No Jet)&KP&0.72&0.01&0.02&3.34\\
&Tribble&0.77&0.01&0.01&3.60\\
3C28&JP&1.21&0.01&0.01&1.35\\
&KP&1.12&0.02&0.02&1.12\\
&Tribble&1.17&0.02&0.01&1.22\\
\hline
\end{tabular}

\vskip 5pt
\begin{minipage}{8.5cm}
Best fitting injection indices for 3C438 and 3C28 and magnetic field strengths. `No Jet' indicates the best fitting injection index values where emission coincident with the assumed location of the jet is excluded (Section \ref{regionselection} and \ref{jvlamodelprms}). Errors are determined using the methods of \citet{avni76} detailed in Section \ref{errors}.
\end{minipage}

\end{table}

\subsection{Model Parameters}
\label{jvlamodelprms}

The plot of $\chi^{2}$ values for varying injection indices are shown in Figure \ref{jvlainjectmin} where one can see that for both sources the minima are model-dependent and occur over a range of values (Table \ref{bestinject}). This is in contrast to our earlier results on other FR-II sources \citep{harwood13} where we used the same method of determining the injection index but found that all models minimized to the same value. The most likely cause of this difference is the use of  truly broad bandwidth observations, providing 29 data points over the frequency space compared to the maximum of 9 used by \citet{harwood13}. As the curvature at GHz frequencies is more tightly constrained, one is able to differentiate between much smaller variations in the models which in turn leads to a variation in the derived injection index values; the observed model dependence is thus not entirely unexpected.

Regardless of the model used, one thing which is immediately clear is that the injection index of both sources remains steeper than the traditionally assumed values of $0.5$ to $0.6$. This is particularly prominent in 3C28 where the injection index is found to be between 1.12$\pm^{0.02}_{0.02}$ (KP) and 1.21$\pm^{0.01}_{0.01}$ (JP), up to double the previously assumed values. Although not as dramatic as those seen in 3C28, the injection index of 3C438 is also found to be relatively steep with values ranging between 0.78$\pm^{0.02}_{0.02}$ (KP) and 0.84$\pm^{0.01}_{0.01}$ (JP) when the whole source is considered. However, unlike 3C28 or those sources investigated by \citet{harwood13}, 3C438 has bright emission coincident with the assumed location of the jet (Figure \ref{combinedimages}) which complicates the determination of the injection index.

\begin{figure}
\centering
\includegraphics[angle=0,width=8.2cm]{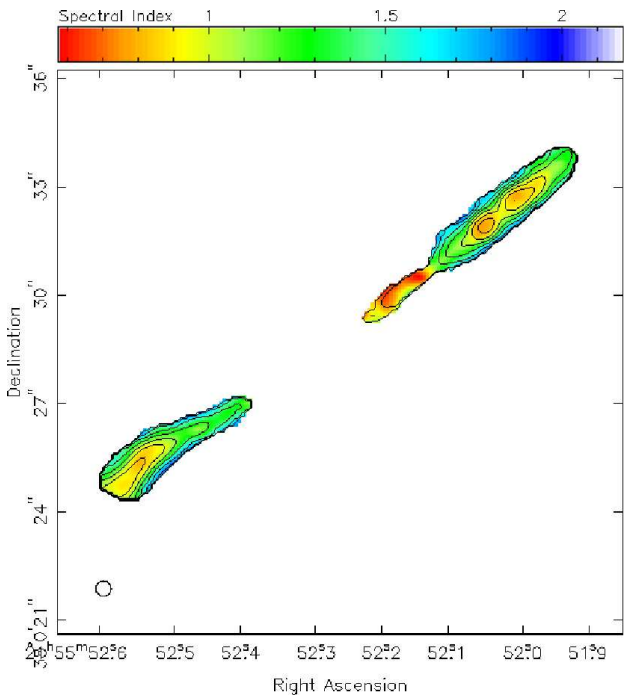}\\
\caption{Spectral index map of emission coincident with the assumed location of the jet in 3C438. The spectral index fitting is performed between 1.4 and 7.9 GHz using the weighted least-squares method described in Section \ref{specindexfitting}.}
\label{jetspectralindex}
\end{figure}

The jets of FR-IIs are still poorly understood and it is therefore not clear whether the plasma responsible for this emission is due to reaccleration of lobe material at a jet-lobe boundary, or if it is accelerated internally within the jet itself. In either case, there is no \emph{a priori} reason to believe that the observed jet emission should be well described by a spectral ageing model. We therefore ran a second minimization with the jet excluded from consideration. From Figure \ref{jvlainjectmin} and Table \ref{bestinject}, one can clearly see that with the jet excluded a minimum occurs at a decreased injection index of between 0.72$\pm^{0.01}_{0.02}$ (KP) and 0.80$\pm^{0.01}_{0.01}$ (JP). At first glance, this reduction in injection index is counter-intuitive as these bright regions are thought to be due to freshly accelerated plasma and so should contain the flattest spectrum emission. However, we see from Figure \ref{jetspectralindex} that a large fraction of the emission has a spectral index of $\alpha > 1$. As these jet regions are also described by the ageing models (even if incorrectly) as having a low spectral age (see Section \ref{jvlaspecage} and Figure \ref{3C438specagemap}) they are not subject to additional curvature at low frequencies and maintain their steep, power law spectra driving the injection index to higher values. Given the non-negligible effect of the jet on the injection index, we perform all subsequent analysis with the jet regions both included and excluded.
 
For both sources, we use the best fitting injection index values of Table \ref{bestinject} for the final model fitting and determination of the equipartition magnetic field strength as detailed in Section \ref{modelfitting}, with an initial electron energy power law index of $\delta = 2\alpha_{inj} + 1$. We find a magnetic field strength in the range of $3.65$ nT (KP) to $4.02$ nT (JP) for 3C438, $3.34$ nT (KP) to $3.77$ nT (JP) for 3C438 where the jet has been excluded and $1.12$ nT (KP) to $1.35$ nT (JP) for 3C28. These magnetic field strengths, summarised in Table \ref{bestinject}, are used for all subsequent analysis.

\subsection{Spectral age and model comparison}
\label{jvlaspecage}

\begin{table*}
\centering
\caption{Model Fitting Results}
\label{jvlarestab}
\begin{tabular}{ccccccccccc}
\hline
\hline

Source&Model&Mean $\chi^{2}$&Mean $\chi^{2}_{red}$&\multicolumn{5}{c}{Confidence Bins}&Rejected&Median\\
&&&&$<$ 68&68 - 90&90 - 95&95 - 99&$\geq$ 99&&Confidence\\

\hline
3C438&JP&47.17&1.75&2229&1214&498&1203&4108&Yes&$>$ 99\\
&KP&45.87&1.70&2416&1228&546&1175&3887&Yes&$>$ 99\\
&Tribble&46.85&1.74&2256&1237&513&1196&4050&Yes&$>$ 99\\
3C438&JP&45.05&1.67&1779&1019&449&956&2930&Yes&$>$ 99\\
(No Jet)&KP&43.16&1.60&2013&1053&472&932&2663&Yes&$>$ 99\\
&Tribble&44.42&1.65&1828&1053&448&951&2853&Yes&$>$ 99\\
3C28&JP&34.45&1.42&1688&307&126&164&644&No&$<$ 68\\
&KP&35.04&1.17&1711&280&90&164&684&No&$<$ 68\\
&Tribble&34.41&1.27&1705&300&108&162&654&No&$<$ 68\\
\hline
\end{tabular}

\vskip 5pt
\begin{minipage}{17.8cm}
`Model' refers to the spectral ageing model fitted to the target listed in the `Source' column. Mean $\chi^{2}$ lists the average $\chi^{2}$ over the entire source with an equivalent reduced value shown in the `Mean $\chi^{2}_{red}$' column. `Confidence Bins' lists the number of regions for which their $\chi^{2}$ values falls with the stated confidence range. `Rejected' lists whether the goodness-of-fit to the source as a whole can be rejected and `Median Confidence' the confidence level at which the model can or cannot be rejected.
\end{minipage}

\end{table*}

\subsubsection{3C438}
\label{3C438specage}

\begin{figure*}
\centering

\includegraphics[angle=0,width=18cm]{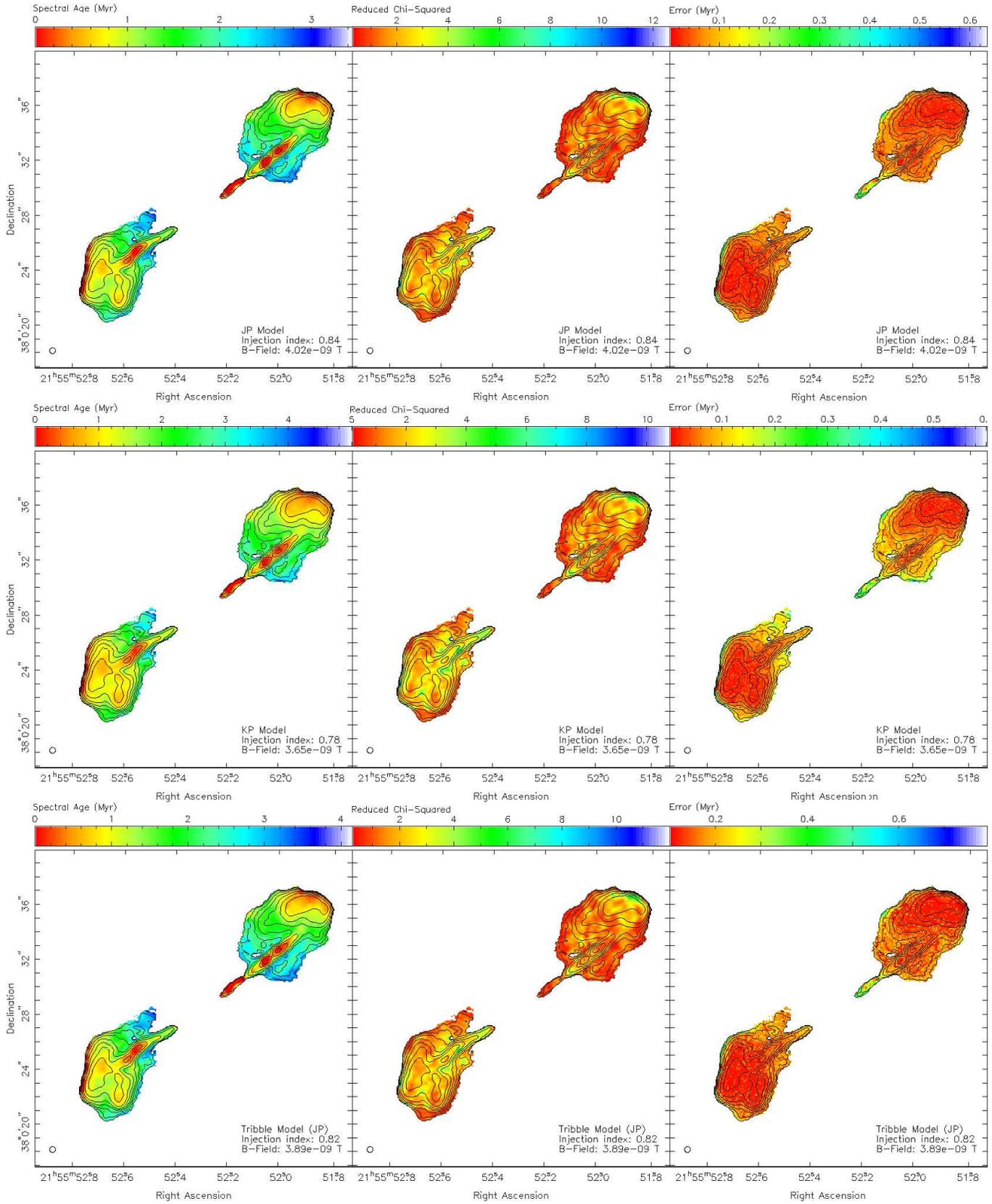}

\caption{Spectral ageing maps (left), corresponding $\chi^{2}$ maps (middle) and error maps (right) of 3C438 with 7.9 GHz flux contours. Three model fits are shown; JP model (top), KP model (middle) and Tribble model (bottom) using the best fitting injection index given in Table \ref{bestinject}.}
\label{3C438specagemap}
\end{figure*}

\begin{figure*}
\centering

\includegraphics[angle=0,width=17.8cm]{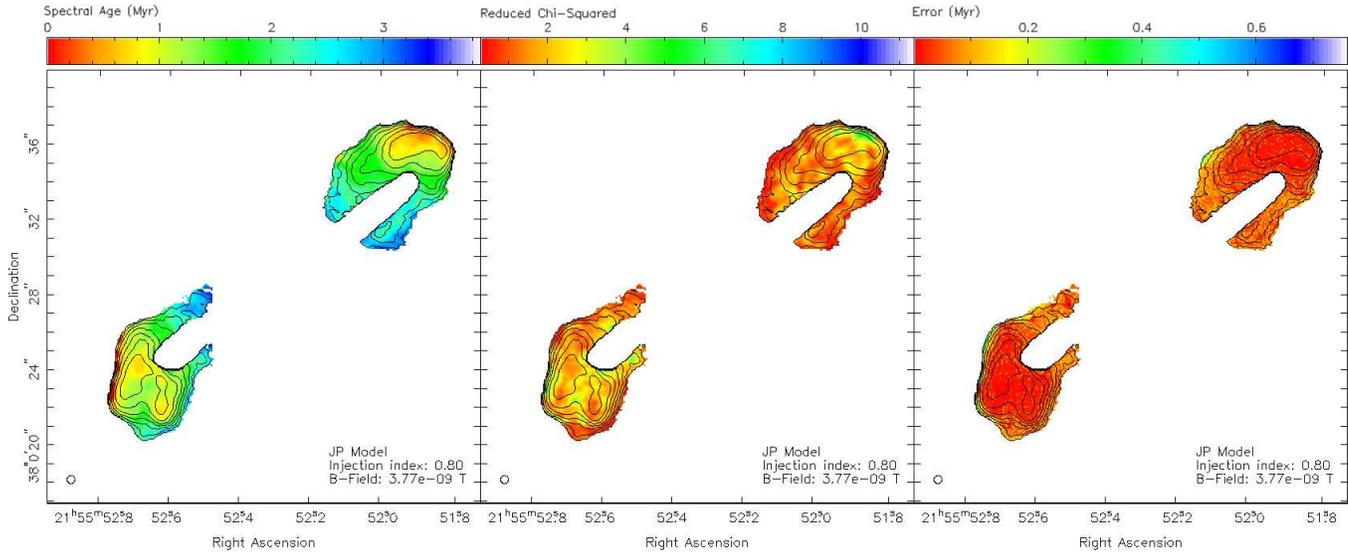}

\caption{Spectral ageing map (left), corresponding $\chi^{2}$ map (middle) and error map (right) of 3C438 with emission coincident with the assumed location of the jet excluded and overlaid with 7.9 GHz flux contours. As all models follow the same spatial distribution of ages shown in Figure \ref{3C438specagemap}, only the JP model fit is shown here for conciseness. The best fitting injection index given in Table \ref{bestinject} is used with the maximum spectral ages of all models given in Table \ref{lobespeed}. }
\label{3C438specagemap_nojet}
\end{figure*}

Figures \ref{3C438specagemap} and \ref{3C438specagemap_nojet} show the spectral age, $\chi^{2}$ and error values as a function of position for the JP, KP and Tribble models for 3C438. We see that, even when the obvious variations due to the jet are ignored, significant cross-lobe age variations exist in a manner consistent with the findings of \citet{harwood13}. Low age regions in the northern and southern lobes are coincident with the hotspots as one would expect, with older regions of plasma residing closer to the core and the main spectral features remaining independent of model type. With the jet regions excluded (Figure \ref{3C438specagemap_nojet}) we see that the spatial distribution of ages mirrors that of the full source as expected. From Table \ref{lobespeed} one sees that for all models the spectral ages have increased in the case where the jet is not considered due to the steeper injection indices and lower magnetic field strengths.

A particularly interesting spectral feature which is observed in Figures \ref{3C438specagemap} and \ref{3C438specagemap_nojet} is the low age region at the edge of the southern lobe. It is not immediately obvious why particle acceleration should be occurring over such an extended region of the source, so further testing is required. We therefore used \textsc{brats} and the L- and X-band presented by \citet{treichel01}, reimaged at 0.42 arcsec resolution, to produce a two point spectral index map. Figure \ref{vlamerlinspecmap} shows the resultant image zoomed to the southern lobe. Although a small region of low spectral index is observed, there is a distinct lack of the extended flat spectrum emission which one would expect if significant particle acceleration was occurring. The $\chi^{2}$ maps also provide some insight into this spectral feature. The small area of flat spectrum / low age seen in both the spectral index and ageing maps is a very poor fit, with the remainder of this region having higher $\chi^{2}$ values compared to the source as a whole. From Figure \ref{combinedimages} one can see a sharp gradient in flux on which this edge lies and it is likely that this low age region is due to the sharp edge of the lobe. We therefore consider the spectral ages derived not to represent the true age of the plasma within this elongated region for the remainder of this paper.

In addition to the lobes, 3C438 has significant emission coincident with the assumed location of the jet. As was discussed in Section \ref{jvlamodelprms}, there is no reason to believe that it should be well described by a spectral ageing model; however, it does provide a good proxy for changes in the steepness of this emission's spectrum. In the southern lobe, changes in the spectrum are observed to be relatively smooth with an increase in brightness and flattening of the spectrum occurring as the jet passes into the observed lobe emission. At this same location, a change in direction of the jet emission is also seen to occur. In the northern lobe, the observed jet emission remains straight along its length, but is seen to be `knotty' with sharp changes in the spectrum. The errors associated with these regions (Figure \ref{3C438specagemap}) are higher than those for the surrounding plasma; however, from the $\chi^{2}$ maps one can see that the bright regions of jet emission are well fitted by spectral ageing models, but are a poor fit in lower flux regions. This is perhaps not surprising as, if the accelerated particles take the form of a power law, it will be well described by a spectral ageing model of zero age, regardless of any subsequent processes. We discuss this jet emission further in Section \ref{jvlajets}.

Although the distribution of ages is consistent with what one would expect from spectral ageing, the goodness-of-fit of the models tested is far less clear. From Figure \ref{3C438specagemap} one can see that the overall $\chi^{2}$ values are fairly consistent over the lobe emission ranging between $\chi^{2}_{reduced} \approx$ 1 and 5 (excluding the southern edge described above). The northern lobe provides a noticeably better fit compared to the southern lobe, which may give a hint towards the cause of these high $\chi^{2}$ values; we discuss this point further in Section \ref{jvlamodelcomp}. The northern hotspot is also poorly fitted by all 3 models of spectral ageing, a feature which was also common in the sources investigated by \citet{harwood13}. The two southern hotspots do not show these high values of $\chi^{2}$ but they are much weaker than their northern counterpart. From Table \ref{jvlarestab} we find that for the goodness-of-fit of the 3 models, KP provides the best fitting model, the JP model providing a noticeably poorer model of the source and the Tribble model providing an intermediate goodness-of-fit. However, all 3 models are rejected at the 99 per cent significance, which given the overall consistency of the fit to the lobes is a somewhat surprising result. We discuss the possible cause of these systematically high $\chi^{2}$ values further in Section \ref{jvlamodelcomp}.

\begin{table*}
\centering
\caption{Lobe Advance Speeds}
\label{lobespeed}
\begin{tabular}{lccccccccc}
\hline
\hline
Source&Model&Lobe&Max Age&+&-&Distance&Speed&+&-\\
&&&(Myrs)&&&(kpc)&($10^{-2}$ $v/c$)&&\\
\hline
3C438&JP&North&3.00&0.05&0.08&30.45&3.313&0.052&0.088\\
&KP&North&4.35&0.40&0.29&30.45&2.285&0.210&0.152\\
&Tribble&North&3.61&0.10&0.07&30.45&2.753&0.076&0.053\\
&JP&South&3.00&0.05&0.08&25.45&2.769&0.046&0.074\\
&KP&South&4.35&0.40&0.29&25.45&1.909&0.176&0.127\\
&Tribble&South&3.61&0.10&0.07&25.45&2.301&0.064&0.045\\
3C438&JP&North&3.36&0.07&0.07&30.45&2.958&0.062&0.062\\
(No Jet)&KP&North&5.50&0.95&0.41&30.45&1.807&0.312&0.135\\
&Tribble&North&4.19&0.10&0.10&30.45&2.372&0.057&0.057\\
&JP&South&3.36&0.07&0.07&25.45&2.472&0.052&0.052\\
&KP&South&5.50&0.95&0.41&25.45&1.510&0.261&0.113\\
&Tribble&South&4.19&0.10&0.10&25.45&1.982&0.047&0.047\\
3C28&JP&North&12.42&0.44&0.44&59.63&1.567&0.056&0.056\\
&KP&North&18.38&1.10&0.84&59.63&1.059&0.063&0.048\\
&Tribble&North&16.82&0.65&0.64&59.63&1.157&0.045&0.044\\
&JP&South&12.42&0.44&0.44&62.86&1.652&0.059&0.059\\
&KP&South&18.38&1.10&0.84&62.86&1.116&0.067&0.051\\
&Tribble&South&16.82&0.65&0.64&62.86&1.220&0.047&0.046\\
\hline
\end{tabular}

\vskip 5pt
\begin{minipage}{17.8cm}
`Model' is the spectral ageing model fitted to the target listed in the `Source' column. `Max Age' is the maximum age of the corresponding `Lobe' in Myrs. `Distance' gives the separation between the oldest aged population in the lobe and the hotspot in kpc (note this is not necessarily the length of the lobe itself). `Speed' lists the derived advance speed as a fraction of the speed of light as detailed in Section \ref{jvlalobespeeds}. Note that the lobe advance speeds of 3C28 contain additional significant uncertainties as detailed in Sections \ref{jvlalobespeeds} and \ref{discussion}.
\end{minipage}

\end{table*}

\subsubsection{3C28}
\label{3C28specage}

Figure \ref{3C28specagemap} shows the spectral age, $\chi^{2}$ and error values as a function of position for the JP, KP and Tribble models for 3C28. While the general distribution of ages is what one would expect for an FR-II source (i.e. a gradient in age from the tip to the core). We again see two zero age regions located at the edge of the source where no obvious source of particle acceleration is present. These non-physical, low age regions have now been observed in all FR-II galaxies since the method's development and are therefore likely an unavoidable consequence of fitting on such small spatial scales in such high dynamic range sources. However, as we are primarily concerned in this paper with the general spatial distribution of ages, overall goodness-of-fit, and the oldest regions of plasma, these non-physical regions have a minimal impact on the analysis and conclusion drawn and so are excluded from subsequent analysis.

Comparing Figure \ref{3C28specagemap} to the spatial distribution of ages in 3C438 it is evident that the gradient of ages is far smoother, lacking any compact structure or jet interaction. There are also no low age regions normally indicative of particle acceleration, with youngest age region of the source on the order of between 6 (JP) and 9 (KP) Myrs with a maximum age of approximately 12 (JP) and 18 (KP) Myrs. Comparing the spectral ages to the flux map of Figure \ref{combinedimages}, this is perhaps not too surprising for the northern lobe where no compact structure is observed; however, the southern lobe contains what has so far been assumed to be an active (albeit weak) hotspot. This lack of obvious low age, particle acceleration regions is a significant difference compared to 3C438 and those sources studied by \citet{harwood13} and may have implications for our current understanding of 3C28's current position in its radio life-cycle. We discuss this further in Section \ref{3C28radiolife}.

From the $\chi^{2}$ maps of Figure \ref{3C28specagemap} we see that 3C28 is well fitted across the majority of the source with reduced $\chi^{2}$ between 1 and 2. One particularly interesting feature is the high $\chi^{2}$ regions in the northern lobe peaking at  $\chi^{2}_{reduced} \approx 8$, similar to the high $\chi^{2}$ values we found for the hotspots of 3C300 and 3C436 in our previous work  \citep{harwood13}. From Table \ref{jvlarestab} we find that the goodness-of-fit of the 3 models follows the same sequence as 3C438 with the KP model providing the best fit, the JP model providing a noticeably poorer fit and the Tribble model providing an intermediate goodness-of-fit. However, in contrast to the results for 3C438, all of the models provide a good description of the observed spectrum as a whole with none of the 3 models being rejected at even the 68 per cent confidence level.

\subsection{Lobe Speeds}
\label{jvlalobespeeds}

To determine the characteristic advance speed of the lobes we use the standard method employed by \citet{alexander87}, as in our previous study \citep{harwood13}. Assuming an electron population accelerated at the hotspot, the advance speed of the lobe can be given by $v_{lobe} = d_{hs}/t_{old}$, where $v_{lobe}$ is the characteristic lobe speed, $d_{hs}$ is the distance to the current location of the hotspot and $t_{old}$ is the age of the oldest region of plasma. Note that for the sources presented within this paper we use the term hotspot to refer to the regions of bright emission at the end of the radio lobes, although we acknowledge that this does not fit the strict definition laid out by, for example, \citet{leahy97}.

The advance speeds for both sources are shown in Table \ref{lobespeed}. For 3C438 and using the best fitting injection index we find an advance speed of between 2.29 (KP) and 3.31 (JP) $\times10^{-2}$ $c$ for the northern lobe and 1.90 (KP) and 2.77 (JP) $\times10^{-2}$ $c$ for the southern lobe. These values decrease slightly where the jet is excluded to between 1.81 (KP) and 2.96 (JP) $\times10^{-2}$ $c$ for the northern lobe and 1.51 (KP) and 2.47 (JP) $\times10^{-2}$ $c$ for the southern lobe. These ages fall well within the expected range compared to previous studies of similar sources  (e.g. \citealp{alexander87}).

For 3C28 we derive lobe advance speeds slower than those found in 3C438 with values between 1.06 (KP) and 1.57 (JP) $\times10^{-2}$ $c$ for the northern lobe and 1.12 (KP) and 1.65 (JP) $\times10^{-2}$ $c$ for the southern lobe. Whilst these values appear reasonable compared to previous studies of FR-II sources, the method used may in this case be significantly underestimating the lobe advance speed. The lack of compact emission and, perhaps most significantly, low age emission in 3C28 mean that both $d_{hs}$ and $t_{old}$ are uncertain. In the northern lobe, the distinct lack of a hotspot means that determining the exact location of the initial particle acceleration is difficult; however, given some compact structure typical of a hotspot is observed in the southern lobe (Figure \ref{combinedimages}) the assumption that the sites of particle acceleration in 3C28 are located near the tip of the lobes (determined here by the peak flux), as is observed in the majority of FR-II sources, is not unreasonable. Therefore whilst the value of $d_{hs}$ is uncertain, it is unlikely to change the advance speed by a large amount.

The validity of $t_{old}$ may have a much larger impact on the derived speeds of 3C28. All of the characteristic advance speeds considered so far (3C438, \citealp{alexander87}, \citealp{harwood13}) have been for sources which are currently active and advancing through the external medium. However, the lack of observed low age emission (Figure \ref{3C28specagemap}) means that this may not be the case for 3C28. Under the assumption that the source is no longer active, the time variable for determining the advance speed becomes  $t_{old} = t_{max} - t_{min}$ where $t_{max}$ and $t_{min}$ are the maximum and minimum observed spectral ages respectively, increasing the characteristic advance speed of 3C28 during the active phase by a factor of around 2. We discuss the possibility of 3C28 being no longer active and the implications of this further in Section \ref{3C28radiolife}.

%% file: discussion.tex
\section{Discussion}
\label{discussion}

Over the last 40 years, models of spectral ageing have become a commonly used tool in determining the age, hence dynamics, of powerful FR-II galaxies. However, as was discussed in Section \ref{specagemodels}, the validity of a number of assumptions used in the application of these models has recently been called into question. When more tightly constrained by modern, broad bandwidth observations, \citet{harwood13} have shown that, for at least two sources, model parameters such as the injection index, and general assumptions such as negligible cross-lobe age variations, are less reliable than previously thought. These findings, combined with an uncertainty as to which spectral ageing model best describes emission from the lobes of FR-II galaxies and the well known spectral vs dynamical age disparity \citep{eilek96a, harwood13}, means that if spectral ageing is to remain a valid tool for the investigation of such sources, it is vital that the cause of these problems are well understood. In this paper, we have presented results which further expand on the sample presented by \citet{harwood13} and, for the first time, tightly constrain curvature at frequencies of a few GHz using truly broad bandwidth observations. In the following sections we relate these new results to previous findings and discuss how this changes our overall understanding of radio galaxy dynamics.

\begin{figure}
\centering
\includegraphics[angle=0,width=8.4cm]{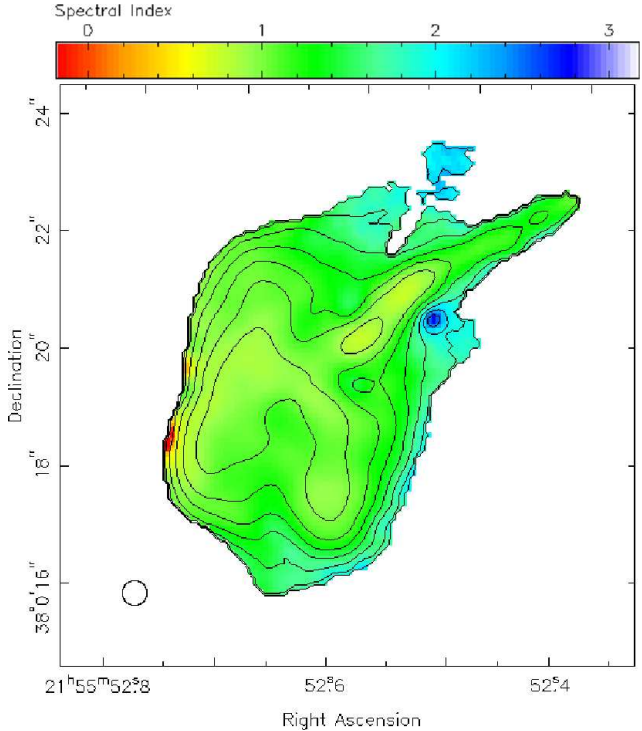}\\
\caption{Two-point spectral index map created in \textsc{brats} of the southern lobe of 3C438 between 1.4 and 8 GHz. The data used is that presented by \citet{treichel01}, reimaged at 0.42 arcsec resolution.}
\label{vlamerlinspecmap}
\end{figure}

\subsection{Injection index}
\label{injectionindex}

One of the key findings of \citet{harwood13} was the deviation of the injection index from previously assumed values of around 0.5 to 0.6. It is again clear from the results presented in Section \ref{jvlamodelprms} that this discrepancy remains, with values ranging between 0.78 (KP) and 0.84 (JP) for 3C438 and between 1.12 (KP) and 1.21 (JP) for 3C28. It was noted in Section \ref{jvlamodelprms} that for 3C438, at least some of this difference is likely to result from the inclusion of emission coincident with the strongly interacting jets. There is no \emph{a priori} reason to believe this emission will be well fitted by models of spectral ageing and so could be forcing the derived injection index to steeper values; however, from Table \ref{bestinject} we see that even with the jet removed the injection index remains $>$ 0.7 for all models and so cannot alone bring these values back into agreement. 

The very steep injection index of up to 1.21 for the JP model of 3C28 presents an even bigger challenge to explain. The most straightforward physical interpretation of such a steep injection index is that the jet of 3C28 terminates, or terminated, in a very weak shock. Assuming a simple, non-relativistic hydrodynamical shock in which the magnetic field of the jet fluid is weak\footnote{In reality, the hotspot may be a relativistic hydrodynamic shock of a highly magnetised jet fluid \citep{kirk00, konar13}, but these assumptions should provide a sufficiently accurate approximation for our purposes.}, the injection index can be related to the Mach number by \citep{blandford87} \begin{equation}\label{machnumber_inject}M = \sqrt{\frac{2\alpha_{inj}+3}{2\alpha_{inj}-1}}\end{equation} which for 3C28 gives a Mach number of between 1.95 (JP) and 2.05 (KP). While this is much weaker than the shock expected from the historically typical injection index value of  $\alpha_{inj} = 0.6$ ($M = 4.6$), lower Mach numbers are not without precedent. For example, a study by \citet{odea09} who used high resolution measurements of hotspots to determine the injection index, found values ranging between $\alpha_{inj} = 0.52$ and $1.11$ for there sample, corresponding to a Mach number of between 2 and 10.

\begin{figure*}
\centering

\includegraphics[angle=0,width=18cm]{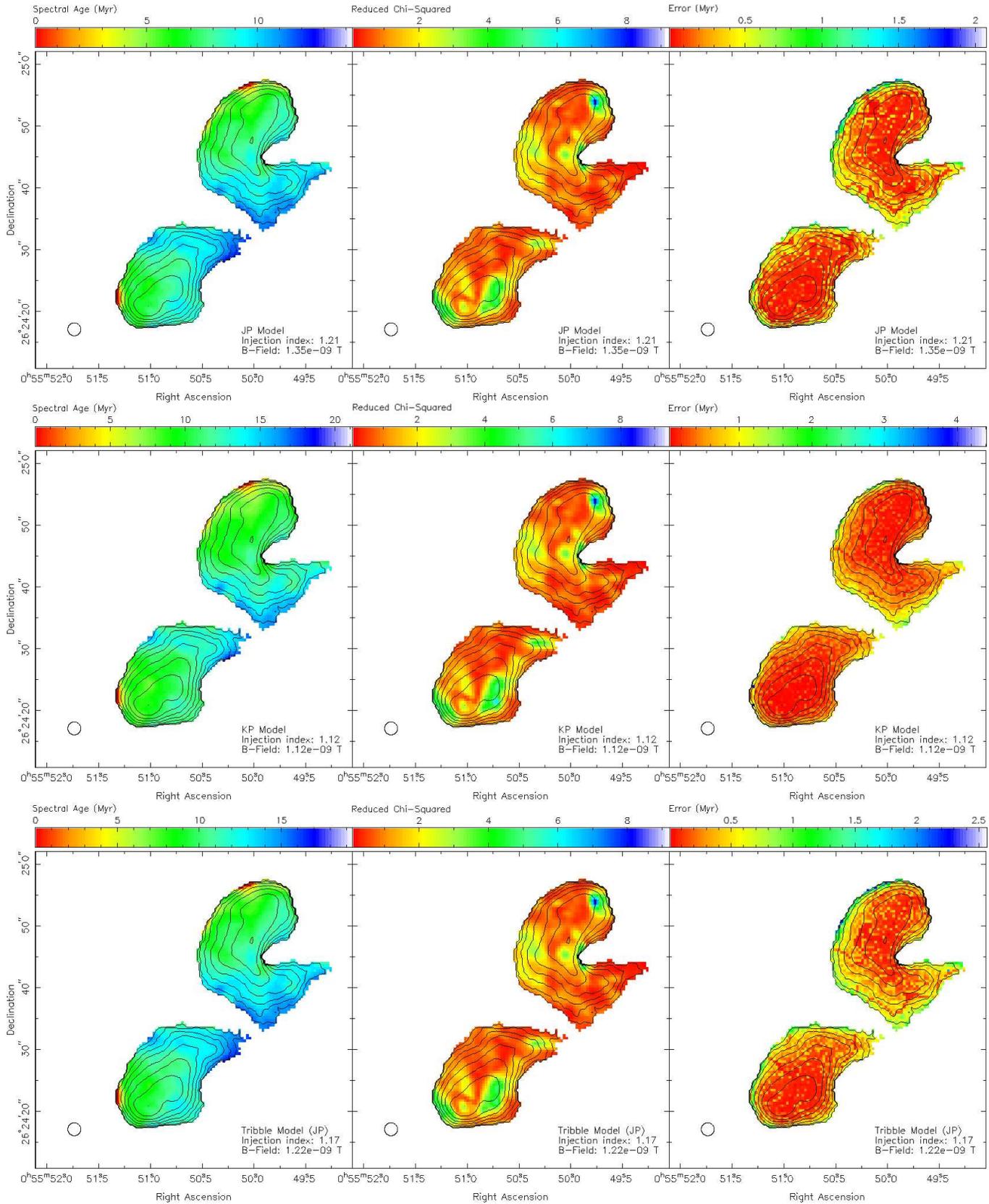}

\caption{Spectral ageing maps (left), corresponding $\chi^{2}$ maps (middle) and error maps (right) of 3C28 with 7.9 GHz flux contours. Three model fits are shown; JP model (top), KP model (middle) and Tribble model (bottom) using the best fitting injection index given in Table \ref{bestinject}.}
\label{3C28specagemap}
\end{figure*}

\begin{figure*}
\centering

\includegraphics[angle=0,width=8.82cm]{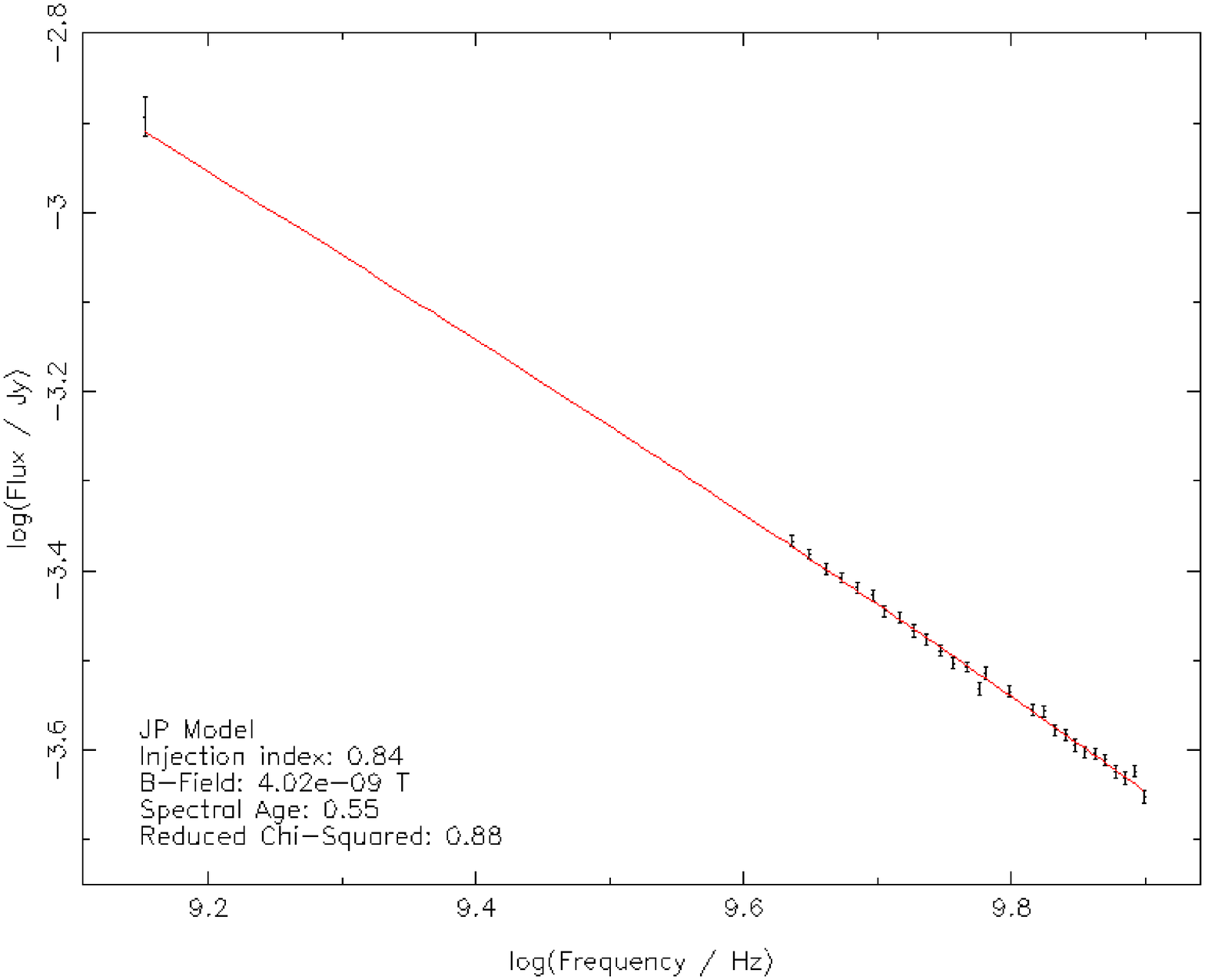}
\includegraphics[angle=0,width=8.82cm]{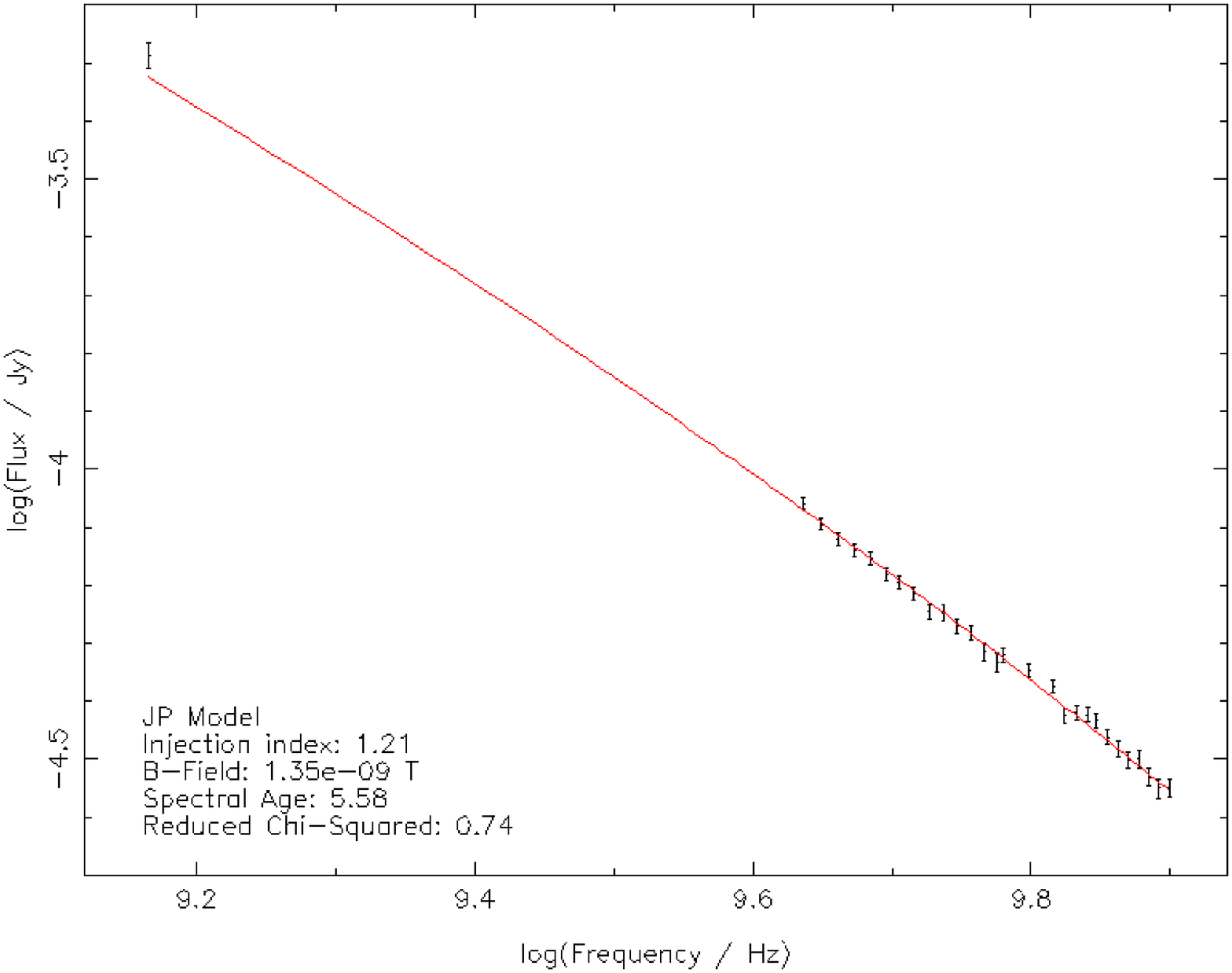}\\
\includegraphics[angle=0,width=8.82cm]{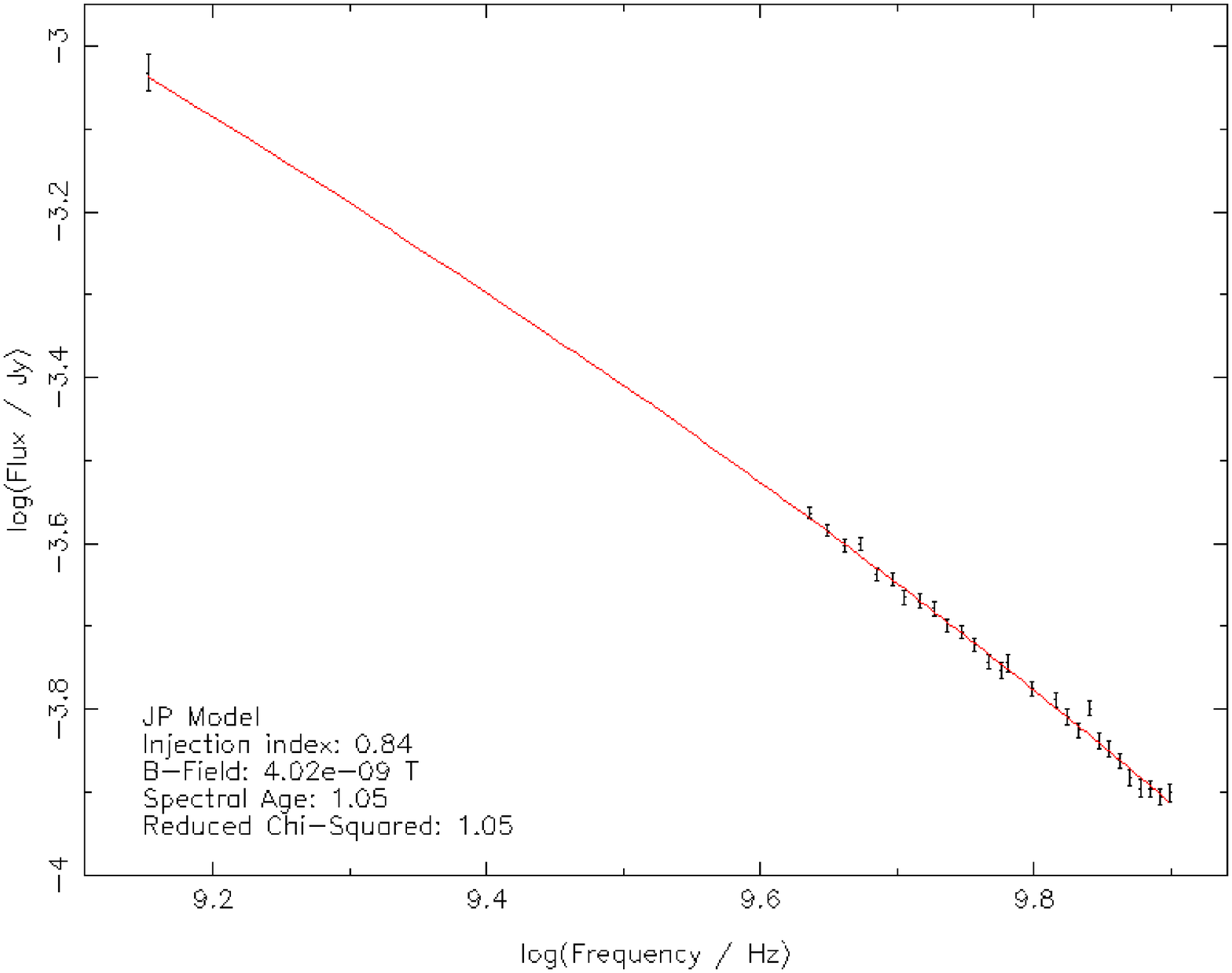}
\includegraphics[angle=0,width=8.82cm]{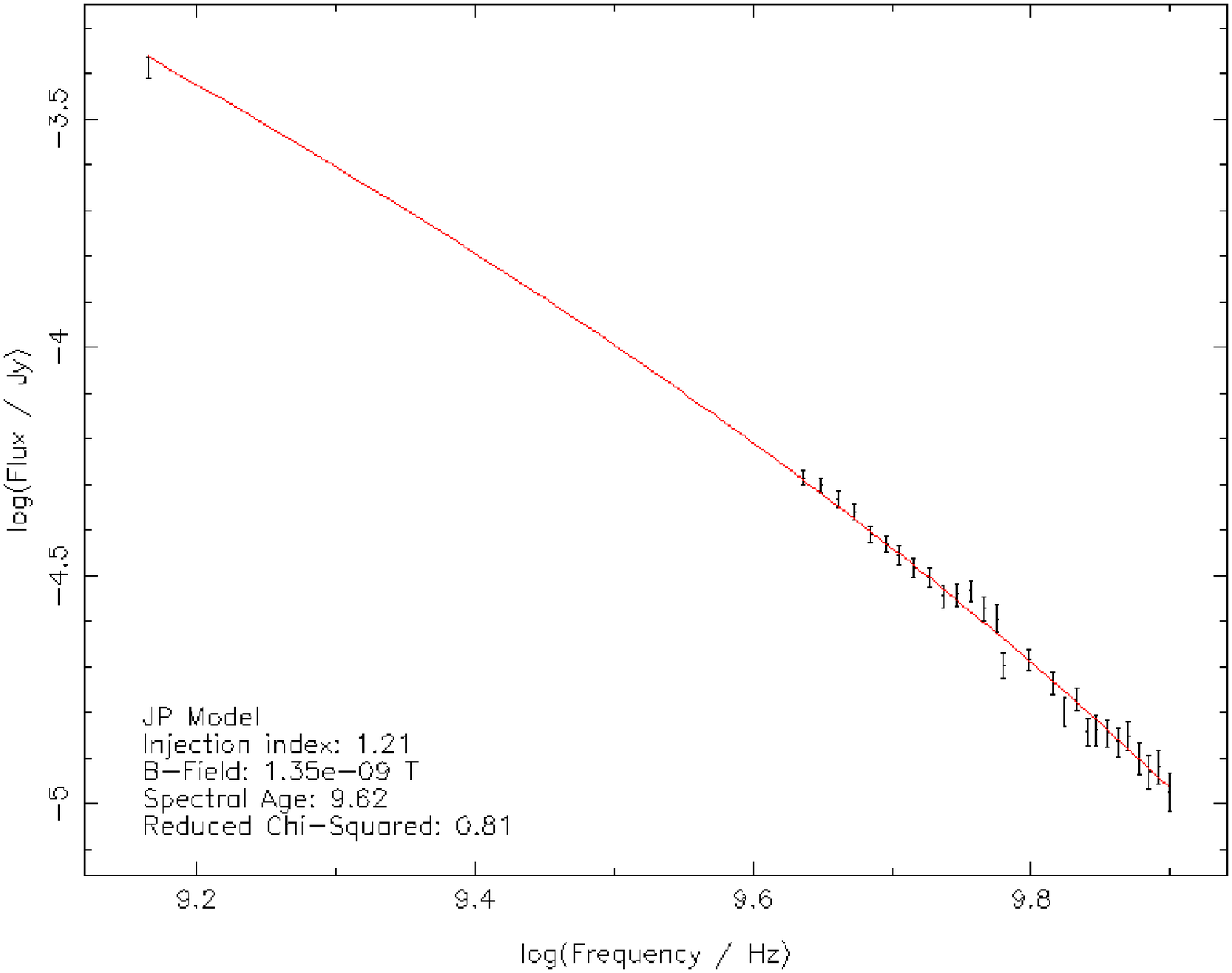}\\
\includegraphics[angle=0,width=8.82cm]{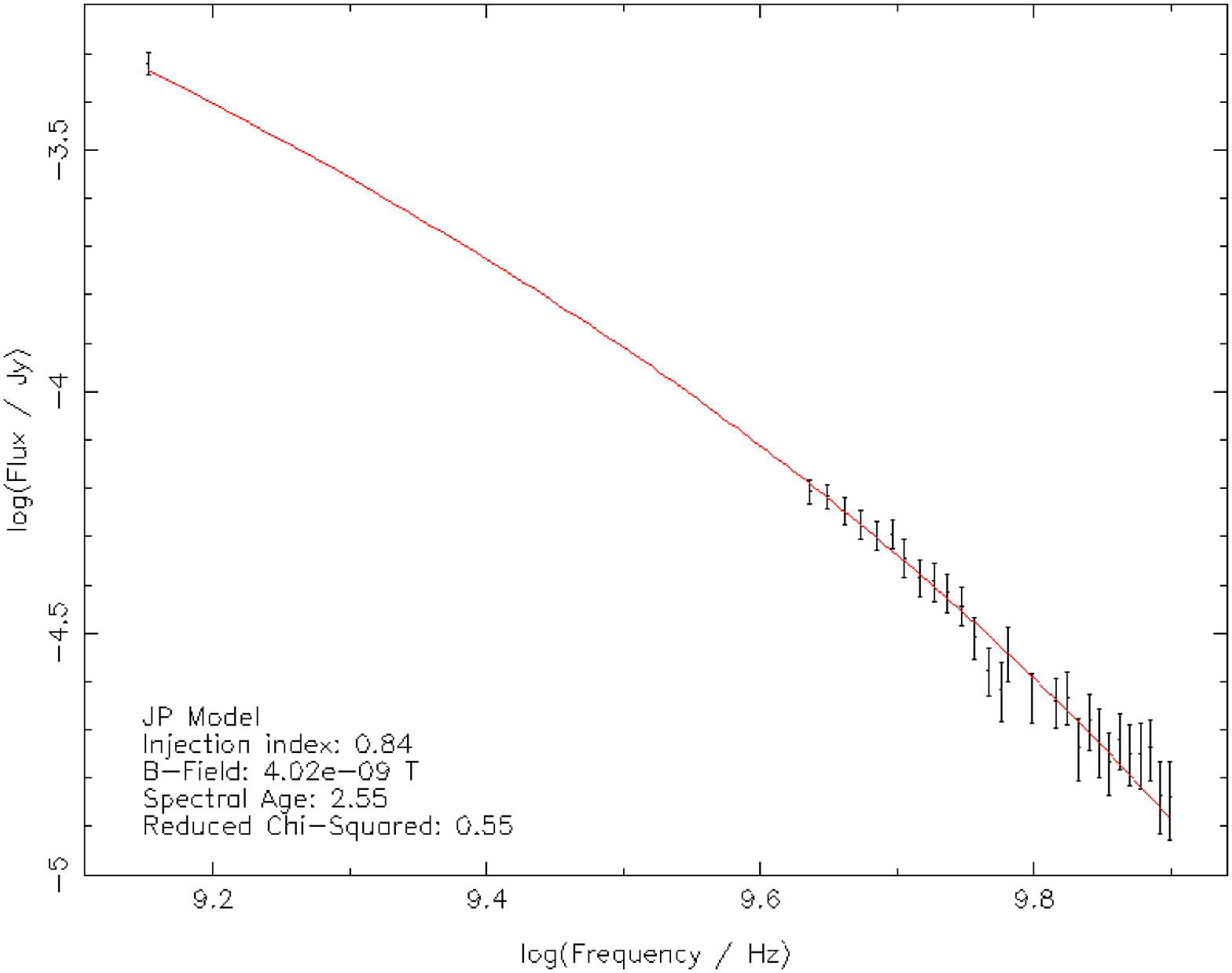}
\includegraphics[angle=0,width=8.82cm]{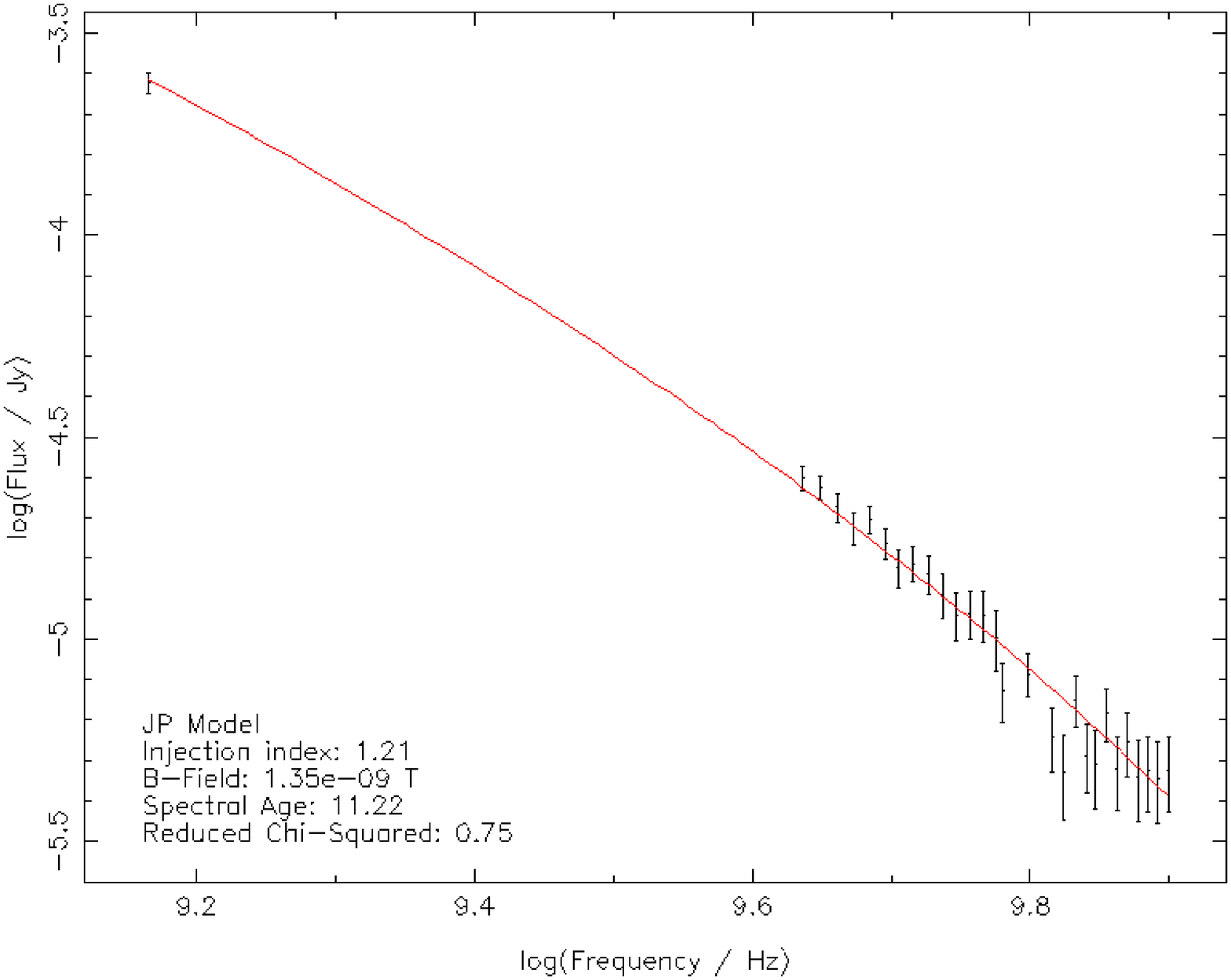}\\

\caption{Plots of flux against frequency of example well fitted regions for 3C438 (left) and 3C28 (right), at low (top), moderate (middle) and high (bottom) spectral ages. Red lines indicate the best-fitting JP model with parameters as noted on each plot.}
\label{jvlagoodfit}
\end{figure*}

An alternative explanation is the reliability of the minimization method used to determine the injection index. Tests on simulated data (e.g. \citealp{stroe14}) have shown that the $\chi^{2}$ minimization used is robust and provides a significant improvement over previous methods such as the use of the spectral index measured in particle acceleration regions; however, while this method is able to use all observed emission in determination of the injection index, rather than just that a small zero age region, its accuracy becomes increasingly model dependent as the fraction of aged plasma increases. The lack of low age emission could therefore be influencing the result if the model is not sufficiently constrained, particularly at lower frequencies. While this may provide some explanation of the very steep injection index observed in 3C28, from Figures \ref{3C28specagemap} and \ref{jvlagoodfit} and Table \ref{jvlarestab} we see that the models do provide a good overall description of the source and are therefore unlikely to account for the full discrepancy.

These findings therefore appear to reinforce the idea that steeper than expected injection indices are widespread. Initial findings from spectral studies of similar FR-II sources using low-frequency (50 - 160 MHz) LOFAR observations (Harwood et. al., in prep) also show that these steeper injection indices are observed at low energies where negligible losses mean that the spectral index of a source should be close to the injection index. Given the lack of strong, distinct hotspots in our observations (and therefore presumably weak jets), the steep injection indices we measure are physically plausible in models such as those presented by \citet{konar13} which suggest that jet power is the primary factor in determining a source's injection index. However, although we can be confident the injection index values observed are intrinsic to the source, rather than a systematic offset, the physical interpretation of why such a disparity exists requires further investigation to determine conclusively.

\subsection{The jets of 3C438}
\label{jvlajets}

As emission coincident with the location of the jets is observable in 3C438, the bias they may cause to the overall study of the lobe emission must be considered. Comparing the statistics of Table \ref{jvlarestab} for 3C438, we find that for the best fitting KP model when the jet is included, approximately 42 per cent of the fits are rejected at $\geq$ 99 percent confidence, and without the jet only around 37 per cent of model fits are rejected at this level. Conversely, only 26 per cent of the regions show a good fit at the $<$ 68 percent confidence level, compared to 28 per cent when the jet is removed. It therefore appears that the inclusion of the jet may give rise to at least some poorly fitted regions of the source, but at a level which cannot explain the rejection of the models as a whole. We discuss this model rejection further rejection in Section \ref{jvlamodelcomp}.

The observed jet emission also provides the opportunity to consider the dynamics of the jets themselves. As noted in Section \ref{jvlamodelprms}, brightening occurs in both the northern and southern jets coincident with the location where it enters the lobe. One possible explanation of such features is therefore an interaction between the lobe material and the jet itself. The exact cause of such an interaction is unclear, but it could plausibly be due to either lobe material being reaccelerated at the jet-lobe boundary, or the onset of turbulence within the jet resulting in localised shock acceleration. The observed (spatial) curvature in the jet found in the southern lobe may possibly be evidence of this interaction; however, due to the presence of two low spectral age hotspots-like structures (one active and one relic, Figures \ref{3C438specagemap} and \ref{3C438specagemap_nojet}\footnote{Note that these hotspot-like structures are also observed in Figure \ref{vlamerlinspecmap}, but are less obvious due to the colour scale being set such as to emphasise the artefacts discussed in Section \ref{jvlaspecage}.}), we suggest that, assuming an FR-II type jet, the cause is more likely due to transverse movement of the jet between these two regions.

An alternative possibility is that the jets are more like those found in FR-Is with the lobes being formed upstream of the particle acceleration regions, similar to the intermediate FR-I/II source Hydra A. There is some evidence to support this in the southern lobe where we observe a flat spectral index (Figure \ref{jetspectralindex}) and low-age emission with a slight gradient away from the core (Figure \ref{3C438specagemap}) which could plausibly be an FR-I type acceleration region responsible for forming the lobes; however, emission from the northern jet does not display this form of structure and is more 'knot-like'  with is a distinct separation between these regions and the low age emission at the end of the lobe where particle acceleration must also be occurring. It is therefore harder to reconcile the northern lobe with a FR-I type jet. It is possible that the jets are asymmetric, as is observed in hybrid morphology sources (HYMORS) \citep{gopal00}, but this would require intrinsically dissimilar jets, environments and/or acceleration mechanisms to conspire to form lobes that are similar in terms of both morphology and the distribution of ages and is therefore unlikely.

While the majority of the observed jet emission appears (at least in projection) to be embedded within the lobe material, the regions closest to the core extend well beyond the visible extent of the lobes and so cannot be interacting with the lobe plasma. The jets must therefore be interacting with their external environment, with the observed emission being caused either by instabilities within the jet and the onset of turbulence (similar to that with the lobe material discussed above), or through a more direct interaction such as the entrainment of thermal plasma. The morphology and spectrum of these jets alone do not allow us to determine which of these two possibilities is the underlying cause of the emission but, when combined with the ageing analysis and source dynamics discussed below, does allow us to provide potentially interesting insights into the underlying physics off 3C438. We discuss this and its potential impact on outstanding questions in the context of the sources dynamics further in Section \ref{jvlaspectralages}.

\subsection{The age and dynamics of 3C438}
\label{jvlaspectralages}

One of the primary concerns raised in the discussion of spectral ageing is the well known discrepancy between a source's dynamical and spectral age (e.g. \citealp{eilek96a, blundell00}). Thanks to X-ray investigations by \citet{kraft07}, the environment in which 3C438 resides is well known and so we are able to constrain the dynamical age. \citet{kraft07} find that the cluster in which 3C438 is located has a core radius of approximately 30 arc seconds, so that the source resides entirely within a region of approximately constant density. The dynamics of the radio source are thus particularly simple (e.g. \citealp{falle91}) \begin{equation}\label{dynamicseq}L = a\left(\frac{Q}{\rho}\right)^{1/5} t^{3/5}\end{equation} where $L$ is the length of the jet, $Q$ is the jet power, $\rho$ is the environment density, $t$ is the time the source has been active and $a$ is a dimensionless constant. We can estimate $a$ by assuming that the expansion of the source is ballistic, and thus relativistic, on scales smaller than the characteristic scale defined by \citet{falle91} \begin{equation}\label{constanteq}l_c = \sqrt{\frac{Q}{\rho c^3}}\end{equation} (here we assume a relativistic jet where the effective mass flux is $Q/c^2$). Then at some time $t_0$ we have ${\rm d}L/{\rm d}t = c$ and $t_0=l_c/c$, which allows us to find $a = 5/3$. In practice the dynamics at the transition between the two phases will be more complicated than this, but the simple analysis should give an approximately correct value of $a$. From the observations of \citet{kraft07} the density can be estimated as $\rho = 5.7\times10^{-23} \rm kg$ $\rm m^{-3}$. To estimate $Q$ we use the relationship of \citet{willott99}, normalized as described by \citet{hardcastle07c}, which gives $Q = 1.3 \times 10^{39}$ W, and we take $L = 50$ kpc. Note that $l_c \approx 30$ pc, so $L \gg l_c$, and the analysis of Falle applies. Finally we have \begin{equation}\label{timeeq}t = \left(\frac{L}{a}\right)^{5/3} \left(\frac{\rho}{Q}\right)^{1/3} \approx 9 {\rm Myrs}\end{equation} Although the value of $Q$ is very uncertain (see \citealp{hardcastle13}) the dependence on $Q$ is weak, so this is reasonably robust, and $t$ is significantly greater than the observed spectral age. We have assumed the source is in the plane of the sky: any projection effects will make $L$ and therefore $t$ larger. A large disparity between the spectral and dynamical ages therefore appears to exist, similar to that observed by \citet{harwood13} even when tighter constraints are placed on the spectral curvature at GHz frequencies by the truly broad bandwidth observations. This age difference is in contrast to the suggestion by \citet{blundell00} who propose that it is possible to reconcile the dynamical age of sources with ages $<10$ Myrs. However, for this to apply it is a requirement that the evolution of the magnetic field over the lifetime of the source is well approximated by the spectral ageing model, which may not be the case for 3C438. Although it is not possible to trace the history of the magnetic field the uncertainty introduced to a low age source is likely to be small compared to those in older radio lobes. While better approximating the magnetic field may therefore play a role in resolving the age disparity it is unlikely to be the only significant factor in resolving this problem.

With the advance speeds determined and the external environment well known, a Mach number can also be derived. The sound speed of the medium through which the lobes are advancing is given by $c_{sound}= \sqrt{\gamma kT / m}$ where $\gamma = 5/3$ is the adiabatic index, $m = 0.5m_{p}$ is the mass of the particles and $T$ is the gas temperature, which in turn one can use to derive the Mach number by the standard equation $M = v/c_{sound}$. \citet{kraft07} find that the cluster environment in which 3C438 resides has a gas temperature of $10.7$ keV (region 2 of \citealp{kraft07}), hence using the advance speeds listed in Table \ref{lobespeed}, we find Mach numbers of between 3.7 [2.9] (KP) and 5.4 [4.8] (JP) for the northern lobe, and 3.1 [2.5] (KP) and 4.5 [4.0] (JP) for the southern lobe where the values in square brackets denote the values with the jet emission excluded. This implies that the lobes are overpressured with respect to the external medium and expanding supersonically. At these high Mach numbers one would expect to observe a strong shock at the advancing edge of the lobes; however, in such a high temperature environment such a shock is unlikely to be observable. The dynamical ages provide a significantly lower advance speed of around $3000$ km s$^{-1}$, equivalent to Mach $\approx$ 1.6. This means that in both cases the lobes are overpressured although, if the dynamical ages are correct, there is no expectation of a strong shock at the leading edge of the lobes. 

An additional way we can investigate the lobe pressures of 3C438 is through its energetics. From the \textsc{synch} code of \citet{hardcastle98} (used in Section \ref{modelfitting} to determine the magnetic field strength), we find the total energy density for each radio lobe is between $8.88 \times 10^{-12}$ (KP, excluding jet) and $1.28 \times 10^{-11}$ (JP, including jet) J m$^{-3}$, with a subsequent internal pressure, given by $P=U/3$ where $U$ is the total energy density, of $2.96$ and $4.28$ $\times 10^{-12}$ Pa. The gas pressure of the external medium can also be derived by the standard equation $P=nkT$ where $n \approx 2.2 n_{0}$ is the number density of particles. Using the gas temperature of $10.7$ keV and number density of protons of $3.4 \times 10^{-2}$ cm$^{-3}$ presented by \citet{kraft07}, we derive the pressure of the external medium to be $1.3 \times 10^{-10}$ Pa. These calculations therefore imply that the lobes are \emph{underpressured} by over an order of magnitude. Such underpressures have also been measured in similar sources such as Cygnus A that are found in rich cluster environment (e.g. \citealp{hardcastle10a}), although the discrepancy is usually smaller making 3C438 the most underpressured FR-II source known to date.

One possible explanation for this difference is the uncertainty in the X-ray measurements. \citet{schellenberger15} show that, particularly at high energies, there is a difference between the temperatures measured by Chandra (used by \citealp{kraft07}) and XMM, although it is not yet clear which of these instruments is correct. Assuming the lower XMM value and the offset between the two instruments reduces the temperature to 10 keV (figure 7 of \citet{schellenberger15}) we find that the pressure only reduces to $1.2 \times 10^{-10}$ Pa, so remains significantly underpressured (we would require a temperature of only around 1 keV to bring the lobes back in to pressure balance). It is therefore highly unlikely that instrumental effects alone can resolve this issue.

A more physical explanation for the disparity is that the lobe energy density is dominated by non-radiating particles. As was discussed in Section \ref{jvlajets}, the most plausible cause of the strong emission coincident with the jets is due to an interaction with the external environment. Such an interaction leads to the possibility that thermal matter is being entrained, similar to what is thought to occur in FR-I type radio galaxies (e.g. \citealp{perucho07, laing08}). As the jets of FR-II galaxies must remain highly relativistic out to large distances, entrainment is thought to be low in most cases (e.g. \citealp{leahy91}); however, 3C438 does not display the strong, compact hotspots that are typical of FR-II galaxies. If for 3C438 the cause of these weak hotspots is due to a relatively slow moving jet, then conditions would be favourable for entrainment to occur. In such a scenario, the assumed minimum energy condition would no longer apply, significantly altering the spectral ages and pressures derived for the source. Such pressure disparities are also observed in other sources where material is plausibly being entrained (e.g. the intermediate FR-I/II source Hydra A, \citealp{hardcastle10a}) and so may be relatively common in lobes sources with weak jets. A slow moving jet is also able to provide a natural solution with respect to the observed initial electron energy distribution as, assuming $\Gamma_{jet} < 10$, a steep injection index is to be expected \citep{kirk00}\footnote{Note that this does not explicitly exclude the possibility of a fast jet for which a steep injection index can also be produced \citep{konar13}}.

Another possible explanation comes from investigations at X-ray energies which have shown that the magnetic field strength of many FR-II sources is not in equipartition as has been so far assumed \citep{croston05}. Values range from between 0.3 and 1.3 $B_{eq}$, with the median around 0.7 $B_{eq}$ which could be causing our determination of the internal lobe pressure to be underestimated. Using the JP model with the jet included (where the internal pressure is highest) and assuming the minimum value of 0.3 $B_{eq}$ of \citet{croston05}, we find that the lobes remain underpressured by an order of magnitude, with the pressure increasing to only $1.98$ $\times 10^{-11}$ Pa.

We also know that a maximum spectral age occurs when $B = B_{CMB} / \sqrt{3}$ \citep{harwood13} where $B_{CMB} = 0.318 (1 + z)^{2}$ nT is the equivalent magnetic field strength of the CMB \citep{hughes91} which, at the redshift of 3C438, occurs when $B = 0.31$ $B_{eq}$. This is close to the value used to derive the maximum pressure and increases the spectral age of 3C438 to $\approx$19 Myrs. Therefore, while departure from equipartition provides a possible explanation for the disparity between the spectral and dynamical ages, it is unlikely alone to provide the solution to the underpressured lobes of 3C438. Determining the cause of such disparities will be vital for future studies if we are to understand the dynamics and energetics of these sources.

\subsection{The classification of 3C28}
\label{3C28classification}

While 3C438 presents a morphology typical of many archetypal FR-IIs, 3C28 has historically been much more difficult to classify. While currently classified in the 3CRR catalogue as an FR-II, its non-standard morphology and lack of strong hotspots mean that investigations have both historically  (e.g. \citealp{macklin83, laing83, hardcastle99, chiaberge99}) and recently (e.g. \citealp{donato04, balmaverde06}) deemed 3C28 to be an FR-I; a classification which also persists in major astronomical databases (e.g. the NASA Extragalactic Database\footnote{https://ned.ipac.caltech.edu/}). As within this paper we present both the most sensitive, high resolution images at GHz frequencies of 3C28 to date, along with for the first time detailed information about the source's spectrum on small spatial scales, we briefly consider which of these classifications is correct.

While at low resolutions the unusual morphology of 3C28 understandably causes problems in terms of classification, from Figure \ref{combinedimages} we see that when well resolved the source's overall morphology closely resembles that of an FR-II. The double lobe structure displays significant edge brightening with only a thin jet or bridge observable near the centre of the source and although the northern lobe lacks any compact structure, the southern lobe contains what closely resembles a hotspot, albeit a weak one. The spatial distribution of spectral age shown in Figure \ref{3C28specagemap} also supports the FR-II case, with the gradient of ages increasing towards the core being indicative that the particles were accelerated at the tip of the lobe, rather than close to the core region as is observed in FR-I sources. We are therefore confident that the FR-II classification which has been assumed throughout this paper is robust.

\subsection{Where in the radio lifecycle is 3C28?}
\label{3C28radiolife}

While as discussed in the previous section the distribution of spectral ages are as expected for an FR-II galaxy, they also raise an interesting question: where exactly in the radio lifecycle is 3C28? Intuitively, one would expect the distribution to be arranged such that zero or low age emission is observed in or close regions of particle acceleration which then ages with distance from the hotspot, as is seen in 3C438 (Figure \ref{3C438specagemap}) and in those sources studied by \citet{harwood13}. However, for 3C28 this is not the case and we observe the minimum age to be on the order of around 6 to 9 Myrs (Figure \ref{3C28specagemap}). One possible explanation for this is the combination of a relatively weak shock and the unavoidable consequence of measuring emission from a range of plasma ages along the line of sight. This superposition of spectra is usually assumed to be negligible but if emission along the line of sight to the hotspot is particularly significant it may dominate and cause no low age regions to be observed. While this provides a plausible explanation for the northern lobe where all of the observed emission is highly diffuse, the presence of the bright compact structure in the southern lobe makes this an unlikely scenario. Emission from the southern hotspot dominates the total flux of the region and therefore, assuming current particle acceleration observed in superposition with older regions of plasma, should still display a reduced age relative to the surrounding plasma, even if that age is non-zero. The smooth distribution of spectral ages across both diffuse and compact structure therefore suggests an alternative scenario.

While the `jet-like' emission close to the centre of the source has previously led to the belief that 3C28 is currently active, the lack of a radio (Figure \ref{combinedimages}) or X-ray core \citep{hardcastle06a} supports the idea that the central AGN is turned off, which provides a natural explanation for the observed smooth spatial distribution of spectral ages. Once the central engine becomes inactive and freshly accelerated particles are no longer supplied via the hotspots, the overall brightness of the lobes will begin to fade due to both radiative and adiabatic loses. As adiabatic losses are independent of frequency only the brightness of the observed emission is affected and not the curvature within its spectrum. If therefore the adiabatic losses remain low in the region of the (now inactive) hotspot, one will observe relatively bright, compact structure that is highly curved such as is found in the southern lobe of 3C28. We therefore suggest that 3C28 is currently inactive with the AGN having shut down between 6 and 9 Myrs ago, and the `jet-like'  emission close to the centre of the source being a result of a bridge similar to those observed in the relaxed double 3C442A \citep{comins91, hardcastle97a}, or jet interaction during a previous episode of AGN activity.

\subsection{Model comparison}
\label{jvlamodelcomp}

Determining which, if any, model of spectral ageing best describes the observed emission from FR-IIs is a key step in producing reliable ages for these sources. The model fits of 3C28 provide good agreement with the observed spectrum with none of the models being rejected at the 68 per cent confidence level (Table \ref{jvlarestab}); however, the goodness-of-fit for 3C438 fares much worse. Given the fairly typical morphology and spectral age distribution of 3C438, one would expect the three models of spectral ageing tested to be well fitted to the observations but are all instead rejected at the 99 per cent confidence level. As was discussed in Section \ref{jvlajets} emission coincident with the jet may cause some bias towards higher $\chi^{2}$ values but is unable to fully account for the model rejections. 

Evidence of another cause of this poor goodness-of-fit lies in the northern hotspot region, where strong emission is observed but the models are poorly fitted. One potential cause of this is a differing injection index in the hotspot and the lobes, as has been observed in other FR-II sources (e.g. \citealp{hardcastle01}). If particle acceleration occurs over an extended region rather than at a single point, the lobe emission may be best described by the electron distribution as it leaves the hotspot, rather than in the hotspot itself where turbulence and the mixing of electron populations is likely to be high. The hotspots in the southern lobes do not display such a reduction in the goodness-of-fit, but given this emission is much weaker than its northern counterpart, this is perhaps not surprising. High resolution studies designed to better understand the workings of FR-II hotspots may therefore be required if one wishes to eventually determine a framework for highly reliable, detailed models of spectral ageing.

Although jet and hotspot emission are likely to cause some bias to high $\chi^{2}$ values, insight into the underlying reason for the overall model rejection can also be found by comparing the goodness-of-fit between the two lobes of 3C438. The extended emission in the northern lobe is generally well described by the spectral ageing models compared to its southern counterpart which overall has much higher $\chi^{2}$ values. The worst fitting regions of extended emission in the southern lobe appear to be located between the two hotspots and close to the path of the jet. As the jet and hotspot sweep across the lobe between the two areas, freshly accelerated particles will become mixed with older regions of plasma. This mixed electron population will not be well described by the single injection models used within this paper and we therefore suggest this to be the primary cause of poor model fits, rather than any inherent problem with the models themselves. Detailed numerical modelling of spectral ages in radio galaxy lobes may be able to reproduce such effects in the future.

An additional point which must be addressed is; which of the three models tested provides the best description of these sources and is such a model physically realistic? From Table \ref{jvlarestab} we see that for both 3C438 and 3C28, the KP model provides the best description of the observed spectra, but an environment in which the pitch angle remains fixed (the characteristic feature of the KP model) requires a set of finely tuned parameters and therefore provides a less likely physical interpretation. Whilst historically sources in which the KP model provides the best description are not unheard of (e.g. \citealp{carilli91}), they were in general thought to be the exception. However, we find that when considered on small spatial scales, all of the FR-II sources tested both within this paper and by \citet{harwood13} show that the KP models provides a systematically better fit than the more physically plausible JP model, suggesting this may be common in the majority of sources. Determining whether these results are due to intrinsically fixed pitch angles, or if improved time-averaged pitch angle models are required, is key to the ability of spectral ageing in providing a more accurate description of the spectra of FR-II sources in the future. The Tribble model is able to provide (at least to some extent) an answer to this problem, giving the physical realism of the JP model, an improved goodness-of-fit and a more advanced description of the magnetic fields within the lobes. However, it is clear from these results that if we are to begin eliminating spectral ageing models at a statistically significant level, one must be both well sampled in frequency space, so as to tightly constrain model parameters, and fitting performed over very wide frequency range where the spectral variations between the 3 models are significant.

%% file: conclusions.tex
\section{Conclusions}
\label{conclusions}

In this paper we have presented a spectral ageing study of two cluster-centre FR-II sources, 3C438 and 3C28. Using broad bandwidth VLA observations to tightly constrain spectral curvature at GHz frequencies, we have investigated their dynamics and energetics and determined which model of spectral ageing provides the best description of these sources. We have looked at whether significant cross-lobe variations are likely to be common in powerful radio sources and if the high injection indices found in our previous work \citep{harwood13} are still observed when tightly constrained by full bandwidth observations. We have investigated whether emission from the jet of 3C438 affects spectral ageing analysis and test whether the well known dynamical versus spectral age disparity still exist in cluster-core FR-IIs. The key points made within this paper are as follows:

\begin{enumerate}
\item We find that the injection index of both sources are higher than previously assumed, consistent with those found in our previous investigations.\\
\item Jet emission causes a bias towards steeper injection index values but, even for the strong jet emission observed in 3C438, does not account for large difference between the commonly assumed values and those found within this paper.\\
\item Significant cross-lobe age variations are again observed in the lobes of 3C438. We suggest that these variations are a common feature of FR-II sources and so should be considered carefully in all future studies.\\
\item We suggest that Tribble model currently provides the most convincing description when both goodness-of-fit and physical plausibility are considered for both 3C438 and 3C28.\\
\item We derive a Mach number for the lobes of 3C438 of between 3.1 and 5.4 (1.6 for the dynamical age), implying that the radio lobes are overpressured with respect to the external medium. However, the source energetics suggest that the radiating particles and magnetic field at equipartition cannot account for the necessary pressure to support the lobes, similar to what is observed in other rich cluster sources (e.g. Cygnus A). A discrepancy of approximately an order of magnitude makes 3C438 the most underpressured FR-II source known to date.\\
\item We find that even when tightly constrained at GHz frequencies, an order of magnitude disparity between the spectral and dynamical ages remains. We show that for 3C438, both departure from equipartition and the presence of non-radiating particles arising from the entrainment of thermal material are both able to provide a possible solution to this problem.\\
\item We confirm that 3C28 is an FR-II, rather than an FR-I as reported in some previous investigations, based on its morphology and spectrum at high resolution.\\
\item We suggest that 3C28 is a relic source with the central AGN turning off around 6 to 9 Myrs ago.

\end{enumerate}

We therefore confirm that many of the assumptions which are made in relation to spectral ageing models should be carefully reevaluated. The underlying physical reason for these disparities, such as the high injection indices, remains an open question which will be addressed in the near future by low-frequency studies of radio lobes and of hotspots in FR-IIs (e.g. Harwood et al., in prep). We conclude that if one is to determine reliable intrinsic ages, and hence the total power of these sources, then it is vital that these outstanding questions are resolved and, possibly, new models which better account for our expanding knowledge of these sources derived.

%% file: acknowledgements.tex
\section{Acknowledgements}
\label{acknowledgements}

We wish to thank the anonymous referee whose constructive suggestions have helped expand our interpretation of the results. JJH wishes to thank the Netherlands Institute for Radio Astronomy (ASTRON) for a postdoctoral fellowship and the University of Hertfordshire and the Science and Technology Facilities Council (STFC) for funding via the Studentship Enhancement Programme (STEP) award. JHC is grateful for support from the Science and Technology Facilities Council under grant ST/J001600/1. This work has made use of the University of Hertfordshire Science and Technology Research Institute high-performance computing facility. This research has made use of the NASA/IPAC Extragalactic Database (NED) which is operated by the Jet Propulsion Laboratory, California Institute of Technology, under contract with the National Aeronautics and Space Administration. We wish to thank staff of the NRAO Jansky Very Large Array of which this work makes heavy use. The National Radio Astronomy Observatory is a facility of the National Science Foundation operated under cooperative agreement by Associated Universities, Inc.

%% file: appendix.tex
\appendix

\section{J00560226+2627287: A tailed source in the field of 3C28}
\label{nat}

\begin{figure}
\centering
\includegraphics[angle=0,width=8.7cm]{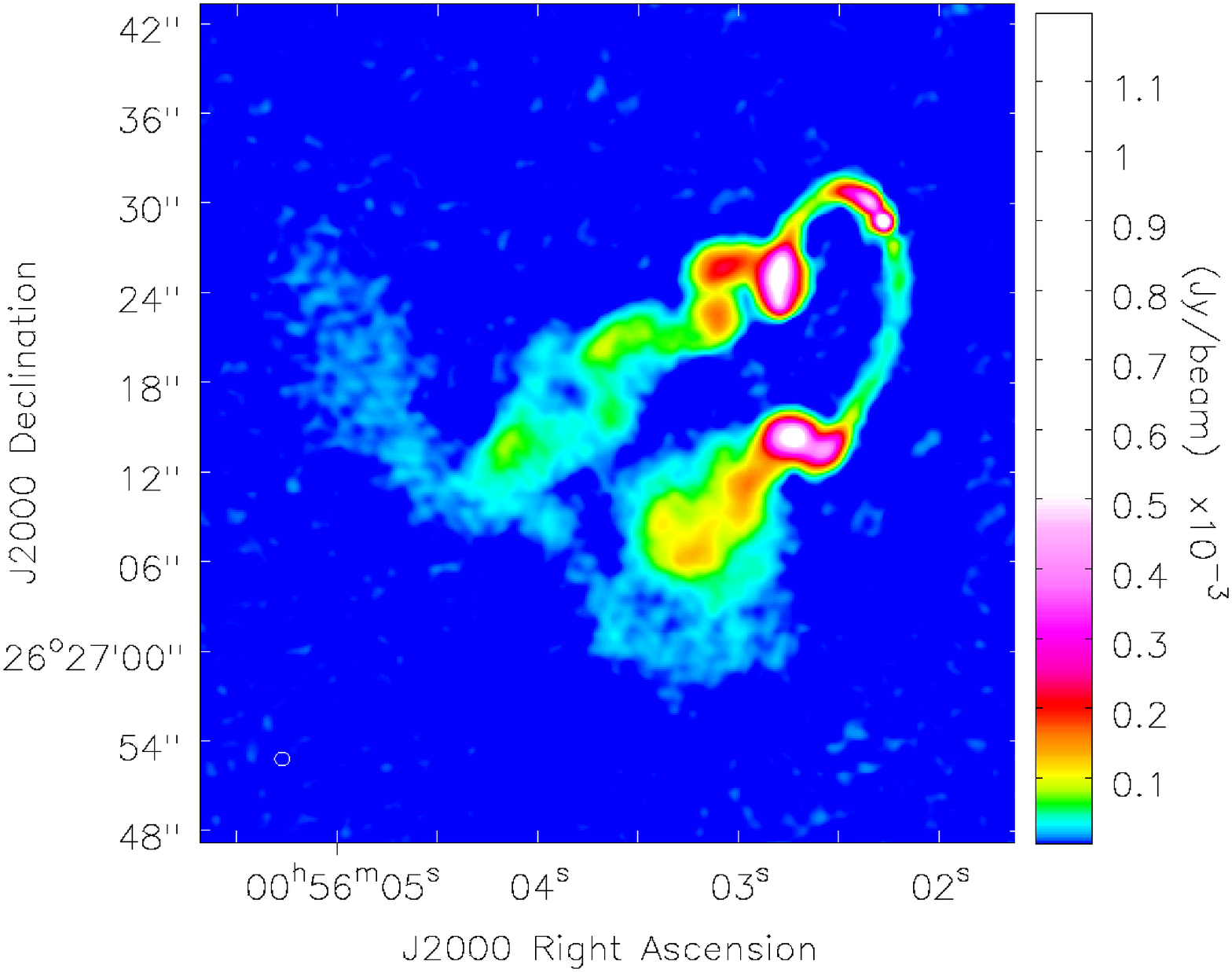}\\
\caption{Combined B- and C-configuration radio map of J00560226+2627287 between 4 and 8 GHz. Imaged using multiscale CLEAN and CASA nterms = 2 to a central frequency of 6 GHz. The off-source RMS of the combined map is 4$\mu$Jy beam$^{-1}$. The restoring beam is 0.99 arcsec.}
\label{combinednat}
\end{figure}

Within the field of view of our observations another large, bright, extended radio source located $\sim$4 arcminutes ($\sim$750 kpc projected) north east of 3C28 was also serendipitously observed (Figure \ref{combinednat}). Unfortunately due to being located at the edge of the JVLA's primary beam and with no well matched L-band data available, we are unable to perform a full spectral analysis; however, we are still able to provide some insight into the source's properties. As this source is located well away from the phase centre, in order to determine reliable flux values a correction for primary beam attenuation was required. Each image was therefore corrected using \textsc{casa}'s IMPBCOR task. As for the JVLA $B_{primary} \approx 45 / \nu$ (where $B_{primary}$ is in arc minutes and $\nu$ is in GHz), at higher frequencies the source is located very close to the edge of the primary beam with some regions falling outside of the corrected image. We therefore restrict our analysis to a frequency range of between 4 and 6 GHz to ensure all of the source is fully encompassed. Integrated flux values were obtained using \textsc{casa} from which we derive a spectral index between 4.3 and 6.0 GHz of $\alpha = 0.98$ using a standard least-squares fit. The integrated flux values and their associated uncertainties are shown in Table \ref{natflux}.

The host galaxy of this source has previously been identified as 2MASX J00560226+2627287 ($z = 0.191399$) and is associated with Abell 115, the same cluster of which 3C28 is a member \citep{beers83, zabludoff90}. From Figure \ref{combinednat} we see that the source contains four major components; a compact core in the north west, a pair of jet-like structures which terminate at bright lobes, and two diffuse tails which, in the plane of the sky, run parallel to the lobes of 3C28. At first sight, the morphology resembles that of a narrow angle tailed radio galaxy (NAT; e.g. \citealp{odea86, odea87}). NATs are nearly always found in rich clusters and thought to be a result of ram pressure bending the jet and lobes as they move through the intracluster medium (ICM), forming their characteristic narrow tails. This was also the conclusion reached in a previous low-resolution radio study of this source by \citet{gregorini89}; however, at the end of the northern tail we also see diffuse emission perpendicular to the main structures which is hard to explain in the context of a NAT as, in the rest frame of the source, it would lie in the flow of material in the external medium. This diffuse feature was also observed by \citet{gregorini89} and so we can be confident that this emission is real, rather than an imaging artefact. The most plausible explanation is therefore that the source is instead a wide-angle tail radio galaxy (WAT; e.g. \citealp{owen76, hardcastle04c}) seen in projection.

\begin{table}
\centering
\caption{Integrated flux for J005603+262717}
\label{natflux}
\begin{tabular}{cccc}
\hline
\hline
Frequency&RMS&Integrated Flux&$\pm$\\
(GHz)&($\mu$Jy)&(mJy)&(mJy)\\
\hline
4.33&33.6&95.4&1.0\\
4.45&36.3&93.2&0.9\\
4.58&37.3&91.9&0.9\\
4.71&32.8&89.8&0.9\\
4.84&35.5&87.7&0.9\\
4.97&34.5&84.9&0.9\\
5.07&38.9&83.9&0.8\\
5.20&33.4&81.3&0.8\\
5.33&38.9&78.8&0.8\\
5.45&38.1&77.5&0.8\\
5.58&39.7&74.7&0.7\\
5.71&36.2&72.8&0.7\\
5.84&38.9&71.5&0.7\\
5.97&43.3&71.0&0.7\\
\hline
\end{tabular}
\vskip 5pt
\begin{minipage}{8.5cm}
Integrated fluxes for J005603+262717 between 4 and 6 GHz. The RMS noise is taken from the image after primary beam correction, in a region away from but representative of the source. The spectral index as determined by a least-squares fit to the values listed above is $\alpha = 0.98$.
\end{minipage}
\end{table}

Similar to NATs, WATs are also observed to reside in rich cluster environments close to the core region, but are thought to be moving more slowly through the ICM resulting in a less pronounced bending of their jets and lobes. \citet{gregorini89} derive a velocity of the host galaxy with respect to the mean cluster velocity of 2400 km s$^{-1}$, much faster than the velocity of $\sim$100 km s$^{-1}$ expected for WATs; however, the minimum physical separation between this source and 3C28 (located at the centre of Abell 115) is $\sim$750 kpc (this is really a lower limit as the redshifts suggests the potential WAT is significantly closer than 3C28). This far away from the centre, the source almost certainly resides in a lower density environment than those WATS observed close to the cluster core, which would allow jets to remain only moderately distorted at these high velocities. 

Another characteristic feature of WATs is that they have so far all been observed to be hosted by massive elliptical galaxies. From measurements made by the 2MASS survey of the host galaxy \citep{skrutskie06}, we see that both the potential WAT and 3C28 (which is known to reside in a massive cD galaxy) have K-band magnitudes of around 13.5, suggesting that the WAT's host galaxy is also similarly massive. This suggests that while WATs have a preference to form in cluster cores due to the abundance of massive elliptical galaxies, it is not a strict requirement. Further investigation is required to substantiate this claim but it provides at least some tentative evidence that WATs in fast moving, massive galaxies further away from the core are also possible given the correct environment.

One further piece of evidence that supports the WAT classification comes from our radio luminosity measurements. WATs are thought to be a subtype of FR-II or transitional FR-I/II radio galaxies. Assuming the spectral index given above and the flux measurements of Table \ref{natflux}, we find a 178 MHz luminosity of $2 \times 10^{26}$ W Hz$^{-1}$, which lies above the FR-I/II break \citep{owen94}. With the northern lobe angled towards the observer, one can also naturally explain the observed sharp turn in the diffuse emission at the end of the northern lobe, the asymmetry between the two jet-like structures, and the northern jet's increased flux density close to the core (due to beaming) through simple geometry. We therefore conclude that the source is most likely a WAT hosted by fast moving, massive elliptical galaxy in the cluster Abell 115.